\documentclass{aastex}
\usepackage{spr-astr-addons}
\usepackage{url}
\usepackage[latin1]{inputenc}
\usepackage{xspace}
\usepackage[french, english]{babel}  
\usepackage{lscape}  
\usepackage{threeparttable} 
\usepackage{dcolumn,amsmath}
\usepackage{graphicx}
\usepackage{bm}
\usepackage{amsfonts}
\usepackage{amsthm}
\usepackage{epsfig}
\usepackage[LGRgreek]{mathastext}
\usepackage[french,refpage]{nomencl}
\usepackage{fancyhdr}
\usepackage{minitoc}
\usepackage{natbib}
\usepackage{multicol}
\usepackage{longtable}

\RequirePackage{color}
\begin{document}
\title{Determination of Cosmological Parameters from Gamma Ray Burst Characteristics and Afterglow Correlations}
\author{H.~Zitouni}
\affil{ PTEAM laboratory, Faculté des sciences, Université Dr
Yahia Fares, P\^{o}le urbain, Médéa, Algeria.}
\email{zitouni.hannachi@gmail.com} \and
\author{N.~Guessoum} \affil{Department of Physics, College of Arts
\& Sciences, American University of Sharjah, UAE.}\email{
nguessoum@aus.edu} \and
\author{W.~J.~Azzam} \affil{Department of Physics, College of
Science, University of Bahrain,
Bahrain.}\email{wjazzam@sci.uob.bh} 

\begin{abstract}
We use the correlation relation between the energy emitted by the
GRBs in their prompt phases and the X-ray afterglow fluxes, in an
effort to constrain cosmological parameters and aiming to
construct a Hubble diagram at high redshifts, i.e. beyond those
found with Type Ia supernovae.

We use a sample of 126 \textit{Swift} GRBs, that we have selected
among more than 800 long bursts observed until April 2015. The
selection is based on a few observational constraints: GRB flux
higher than 0.4 $photons/cm^2/s$ in the band 15-150 keV; spectrum
fitted with simple power law; redshift accurately known and given;
and X-ray afterglow observed and flux measured.

The statistical method of maximum likelihood is then used to
determine the best cosmological parameters ($\Omega_M$,
$\Omega_{\Lambda}$) that give the best correlation for two
relations: a) the Amati relation (between intrinsic spectral peak
energy $E_{p,i}$ and the equivalent isotropic energy); b) the
Dainotti relation, namely between the X-ray afterglow luminosity
$L_X$ and the break time $T_a$, which is observed in the X-ray
flux FX.

 Although the number of GRBs with high redshifts is
rather small, and despite the notable dispersion found in the
data, the results we have obtained are quite encouraging and
promising. The results obtained using the Amati relation are close
to those obtained using the  Type Ia supernovae, and they appear
to indicate a universe dominated by dark energy. However, those
obtained with the correlation between the break time and the X-ray
afterglow luminosity  is consistent with the findings of the WMAP
study of the cosmic microwave background radiation, and they seem
to indicate a de Sitter-Einstein universe dominated by matter.

\end{abstract}


\section{Introduction}
 Gamma-ray bursts (GRBs) are the most powerful explosions in the universe,
 and they occur in galaxies that can be at the farthest reaches of the (observable) universe.
 They have so far been observed with redshifts up to \textit{z} = 9.4, with hopes of reaching \textit{z} = 20 with
 future satellite detectors, qualifying them as potential cosmological probes. Although GRBs
 are not standard candles, the discovery of several luminosity and energy correlations opened
 a new window of investigation in which GRBs could be used to probe cosmological models and
 cosmological issues, like the star formation rate.

    Two GRB luminosity correlations were discovered in 2000.
    The first is a correlation between a burst's luminosity and the time lag between
    the arrival of hard and soft photons in the burst \citep{norris:00}. The second
    is a correlation between a burst's luminosity and its ``spikiness" or variability \citep{fenimore:00}.
    These two correlations were then used to create a GRB Hubble diagram \citep{schaefer:03}. Other luminosity and energy
    correlations were soon discovered, such as: the Amati
    relation \citep{amati:02,{amati:06}, {amati:08}, {amati:09}} which relates the intrinsic spectral peak
    energy, $E_{p,i}$, to the equivalent isotropic energy, $E_{iso}$;
    the Yonetoku relation \citep{yonetoku:04} which is a correlation between $E_{p,i}$ and the isotropic
    peak luminosity, $L_{iso}$; the Ghirlanda relation \citep{{ghirlanda:04},{ghirlanda:06},{ghirlanda:10}} which is a
    correlation between $E_{p,i}$  and the collimation corrected energy $E_{\gamma}$ ; and the Liang-Zhang relation \citep{liang:05}
    which relates $E_{iso}$ to both $E_{p,i}$ and the break time
    of the optical afterglow light curves, $T_a$.

    Early attempts to use GRB correlations to constrain cosmological parameters,
    such as the matter density parameter  $\Omega_M$, faced several problems \citep{{dai:04},{azzam:06},{azzam:06b}}.
    The first problem was the scatter in the correlations, which made it difficult to pin down the cosmological parameters.
    The second was the circularity problem, which refers to the fact that in order to calibrate the luminosity and energy
    correlations, one must assume a cosmological model in the first place. The third problem was the paucity of data points
     with the required input parameters, like the redshift and the spectral peak energy. For a detailed discussion of these issues, the reader is referred
     to the recent review by \cite{amati:13} and the detailed study by \cite{dainotti:13b} who clearly demonstrate the importance of taking proper
     account of the circularity problem, which could, otherwise, affect the evaluation of the cosmological parameters by 10 to 13\%.

    In recent years, a revived interest in GRB cosmology has taken place, perhaps due to the
    new abundance of high quality data. For instance, \cite{petrosian:15} investigated the cosmological evolution
    of a sample of 200 \textit{Swift}  bursts and utilized the results they obtained to put constraints on the star formation rate.
    Another recent study \citep{{wang:15},{wang:16}} provides a thorough investigation of how GRBs can be employed
    to constrain cosmological parameters, dark energy, the star formation rate, the pre-galactic metal enrichment,
    and the first stars \citep{{totani:97},{wijers:98},{mao:98},{mao:10},{porciani:01},{natarajan:05},{hopkins:06},{jakobsson:06},{dermer:07},{daigne:07},{coward:07},{yuksel:07},{kistler:08},{dainotti:15b}}.

    In this paper, we use a sample of 126 Swift GRBs to investigate the correlation
    between the energy emitted by GRBs in their prompt phase and their X-ray afterglow fluxes.
    GRB luminosity correlations necessarily include cosmological parameters through the dependence
    of the luminosity on the burst's distance. The goal is to utilize the correlation between
    burst parameters to infer the best distance function, thus constraining the cosmological
    parameters and creating a Hubble diagram at redshifts that go beyond those found in Type Ia supernovae.

\section{Data Preparation}

The data for the prompt gamma portion was collected fromthe two
official NASA/\textit{Swift}
websites\footnote{http://swift.gsfc.nasa.gov/archive/grb$\_$table/},
\footnote{http://gcn.gsfc.nasa.gov/swift$\_$gnd\_ana.html}. For
the X-ray part of each burst, we used the data published by the
\textit{Swift} public
website\footnote{http://www.swift.ac.uk/xrt$\_$live$\_$cat/}
{\citep{evans:09}.

Until 25.04.2015, \textit{Swift}/BAT had observed 304 GRBs  with
determined redshift. These bursts include 184 ones with an X-ray
counterpart observed by \textit{Swift}/XRT. We eliminated two
GRBs, GRB060708 ($z<2.3$) and GRB090814 ($0.696\leq z \leq 2.2$)
because their redshifts are given very approximately. For
consistency in our calculations, we only keep bursts that have a
spectrum expressed by a single power law (PL) \citep{dainotti:16}
and whose spectral index is given by the \textit{Swift} public
website\footnote{http://swift.gsfc.nasa.gov/archive/grb table/}.
According to  \citep{sakamoto:11} the rule of $\delta_{\chi^2} =
\chi^2_{PL}-\chi^2_{CPL} < 6$ means that the CPL does not improve
significantly the fit, thus the PL can be chosen as an equally
good fit.

With this constraint we are left with 152 GRBs. We then add two
constraints: a duration longer  than two seconds, to keep only
long bursts, and fluxes greater than $0.4~ \mathrm{ph.
cm^{-2}.s^{-1}}$ \citep{ghirlanda:15}. Two other GRBs were
eliminated due to lack of data on their X-ray fluxes, GRB120714B,
GRB080330, and a third a third one, GRB131103A, because of the
presence of a very intense flare at about 1000 s. In the end, we
are left with a first sample of 139 GRBs.

\section{Statistical correlation methods}

We use the maximum likelihood method as described in
\citep{{dagostini:05},{amati:08},{dainotti:13},{dainotti:16}} to
determine correlation relations. The objective is to determine the
parameters $ m $ and $ q $ when interpolating a set of $ N $ data
points $(x_i, y_i) $ by a straight line $y = m~x~+ q$ with
standard deviations $\sigma_{x}$ and $\sigma_y$. Hence, in order
to determine the parameters $(m,q,\sigma_{int})$, we follow the
Bayesian approach of \cite{dagostini:05} by maximizing the
likelihood function $\mathcal{L}(m,q,\sigma)=
exp[-L(m,q,\sigma_{int})]$, such that
\begin{eqnarray}
L(m,q,\sigma)=\nonumber \\
\frac{1}{2}\sum_{i=1}^N \bigg{[}~ln{({\sigma_y^2+m^2\sigma_x^2
+\sigma_{y,i}^2+m^2\sigma_{x,i}^2})}\nonumber\\
+\frac{(y_i-m
x_i-q)^2}{\sigma_y^2+m^2\sigma_x^2+\sigma_{y,i}^2+m^2\sigma_{x,i}^2}\bigg{]},
\end{eqnarray}
 where $\sigma = \sqrt{\sigma_y^2+m^2\sigma_x^2+\sigma_{y,i}^2+m^2\sigma_{x,i}^2}$.
$\sigma_x$, $\sigma_y$, $\sigma_{y,i}$, and $\sigma_{x,i}$
represent the standard deviations corresponding respectively to x,
y, $x_i$ and $y_i$. Because the errors on x and y were unknown,
\cite{dagostini:05} choose $\sigma_x = 0$ and set $\sigma_y =
\sigma_v$. This choice is justified by the fact that y depends on
several hidden parameters, while x does not depend on any factor.
This has recently been used by \cite{wang:16} to justify the
choice of $x = E_p$ and $y = E_{iso}$ instead of the reverse,
because $E_{iso}$ depends on the cosmological parameters.

In this work, since we know neither $\sigma_x$ nor $\sigma_y$ we
replace  $m^2\sigma_x^2+\sigma_y^2$ by  $ \sigma_{int}$, noting
that the latter is called extrinsic scatter by
\citep{{amati:08},{wang:16}} and ``intrinsic scatter" by
\citep{{dainotti:13},{dainotti:16}}. It replaces the sum of all
Gaussian errors that may affect ($x,y$) and which might come from
other non-observed variabilities.

From the expression of $\sigma_{int}$, it is clear that the latter
depends on the slope \textit{m} times $\sigma_{int}$. but
practically speaking, in the minimization procedure for the
function $-\ln{\mathcal{L}}$ with respect to \textit{m} and
$\sigma_{int}$, the corresponding best value of $\sigma_{int}$
must be of the same order as
$\sqrt{\sigma_{y,i}^2+m^2\sigma_{x,i}^2}$. Hence, it depends on
\textit{m} as long as $\sigma_{y,i}$ is not dominant. In this
regard, and in the aim of minimization its value, we choose the
dependent variable $y = f(x)$, which gives a slope \textit{m} such
that $|m| < 1$.

We may refer to the work of \cite{dainotti:15b} in order to
compare the values of $\sigma_{int}$ obtained for different values
of \textit{m} and given in the two tables. In our work and our
sample, with $y = L_X$ and $X = T_a$, we obtain $m = -1.4$ and
$\sigma_{int} = 0.6$. However, if we take $y = T_a$ and $x = L_X$,
we get $m= -0.46$ and $\sigma_{int} = 0.29$. Since smaller values
of $m$ result in smaller values of $\sigma_{int}$, we choose $L_X$
as the $x$ variable and $T_a$ as the $y$ variable. This may be one
of the reasons for the choices made by \cite{amati:03} and
\cite{ghirlanda:04} in their correlations.

We note that the maximization of likelihood function is performed
on the two parameters $(m, \sigma_{int})$, because the parameter
$q$, called ``intercept" is obtained analytically from:
\begin{eqnarray}
q&=&\bigg{[} \sum
\frac{y_i-m~x_i}{\sigma_{int}^2+\sigma_{y,i}^2+m^2\sigma_{x,i}^2}\bigg{]} \nonumber\\
&\times& \bigg{[}\sum
\frac{1}{\sigma_{int}^2+\sigma_{y,i}^2+m^2\sigma_{x,i}^2}\bigg{]}^{-1},
\end{eqnarray}
for each pair $(m,\sigma_{int})$.

For comparison, and in simple cases, we shall use the $\chi^2$ statistical method, which
is defined as follows:
\begin{equation}
\chi^2 = \sum_{i=1}^{N}\frac{(y_i-m~x_i-q)^2}{\sigma^2},
\end{equation}
where $\sigma=\sqrt{\sigma_{int}^2+\sigma_{y,i}^2+m\sigma_{x,i}^2}$.

This method has also been used by \cite{amati:13} to constrain the cosmological parameters. It gives the same results as the maximum likelihood method when the number of data points is large. Statistically, the maximum likelihood method is more reliable when the data set is small \citep{{saporta2011probabilites},{martin2012statistics}}.

\section{Calculation of the energy $E_{iso,\gamma}$ of the prompt gamma emission}

We calculate the total isotropic energy $E_{iso,\gamma}$ emitted by the burst during the prompt phase using the following expression \citep{cardone:11}
\begin{equation}
E_{iso,\gamma}= 4~\pi D_{L}^2(z)~S_{b}~(1+z)^{-1},\label{eq.1b}
\end{equation}
where $S_{b}$ is the bolometric fluence. This quantity is related
to the observed one by \citep{schaefer:07}:
\begin{equation}
S_{b}= S~\frac{\int^{10^4/(1+z)}_{1/(1+z)}E~\Phi_S(E)dE}{\int^{E_{max}}_{E_{min}}E~\Phi_S(E)dE},
\end{equation}
where S$(erg.cm^{-2})$ is the observed quantity corresponding to
the fluence. The integral represents  a correction term
\citep{zitouni:14}. $\Phi_S(E)$ is the mean spectral energy.
$(E_{min}, E_{max})$ is the energy range corresponding to the
observing instrument. The energy range of \textit{Swift}/BAT is
(15 keV, 150 keV). $D_L$ is the GRB luminosity distance computed
in terms of the redshift z,
\begin{eqnarray}
D_L(z)&=&\frac{(1+z)c}{H_0 \sqrt{|\Omega_k|}}
sinn\{\sqrt{|\Omega_k|}F(z)\},\\
F(z)&=& \int^z_0 \frac{dz'}{\sqrt{(1+z')^2(1+\Omega_M
z')-z'(2+z')\Omega_{\Lambda}}},\nonumber
\end{eqnarray}
assuming a standard cosmological $\wedge CDM$ model  with
$\Omega_k=\Omega_m+\Omega_{\Lambda}-1$ , neglecting the radiation
density given by the parameter $\Omega_r$. $c$ is the speed of
light, $H_0$ is the Hubble constant at the present time. The
$sinn$ function is the \textit{sin} function if $\Omega_k > 0$,
corresponding to a ``closed universe", and the \textit{sinh}
function if $\Omega_k < 0$, corresponding to an ``open universe".
For a flat space, $\Omega_k = 0$, thus $D_L$ simplifies to:
\begin{equation}
D_L(z)=\frac{(1+z)c}{H_0}~\int_0^z \frac{dz'}{\sqrt{\Omega_M(1+z')^3+\Omega_L}}.
\end{equation}

\section{Calculation of the luminosity $L_X(t)$ and the energy $E_{iso,XA}$ of the X-ray afterglow}

The luminosity of the afterglow in the X-ray band, $L_X(t)$,
corresponding to a time $t$ measured from the detection of the
prompt emission, is given by the following expression
\citep{sultana:12}:
\begin{equation}
L_X(t)=4~\pi~D_L^2~F_X(0.3-10~keV,t)\times K_c,
\end{equation}
where $F_X(0.3-10~keV,t)$ is the flux observed at the time $t$ in
the X-ray band. $K_c$ is the K-correction for the spectral power
law obtained for each afterglow
\citep{{evans:09},{dainotti:10},{d'avanzo:12},{dainotti:16}}:
\begin{equation}
K_c=\frac{(\frac{10}{1+z})^{2-\Gamma}-(\frac{2}{1+z})^{2-\Gamma}}{10^{2-\Gamma}-0.3^{2-\Gamma}},
\end{equation}
where $\Gamma$ is the measured spectral index. The X-ray afterglow
energy is calculated by integrating over time from its first
detection to its end.
\begin{equation}
E_{iso,XA}=\frac{4\pi D_{L}^2(z)}{1+z}\times
K_c\int_{t_{1}}^{t_{2}}F_X(t) ~dt,\label{eq.2}
\end{equation}
where $t_1$ and $t_2$ are the start and end times of the X-ray
afterglow.

The X-ray afterglow luminosity corresponding to a time $t$ is
calculated from the observed flux at that time. In the
\textit{Swift}/XRT data, we find the X-ray afterglows observed at
the start denoted by $F_{X,early}~~(erg/cm^2/s)$, those observed
eleven hours later denoted by $F_{X,11}$, and those observed after
24 hours denoted by $F_{X,24}$.

\section{Study of the different correlation relations}

In Figure (\ref{fig1azzam3}) we have plotted the energy
$E_{iso,X}$ emitted by the X-ray afterglow against the luminosity
$L_{X,early}$. $E_{iso,X}$ was calculated for a flat universe
 ($\Omega_k=0$, $\Omega_{\Lambda=0.7}$, $\Omega_M=0.3$, $H_0=70~km/s/Mpc$).
 Using the $\chi^2$ statistical method, we note that there is a good correlation between the
 two quantities, except for 13 bursts falling outside of the group: GRB060926, GRB060904B, GRB061110B,
 GRB070411, GRB070611, GRB070506, GRB080411, GRB090529, GRB090726, GRB091024, GRB100902A, GRB120909A, GRB140114A.
 This group is characterized by X-ray afterglow fluxes that either do not have breaks in their temporal
 profiles or have a hint of a break in a highly slanted plateau.

In the rest of our study, we thus limit our sample to 126 GRBs
(the previous 139 minus these 13), as there is a good chance of
finding a strong correlation for them.

\begin{figure}[h]
\centering
\includegraphics[angle=0, width=0.483\textwidth]{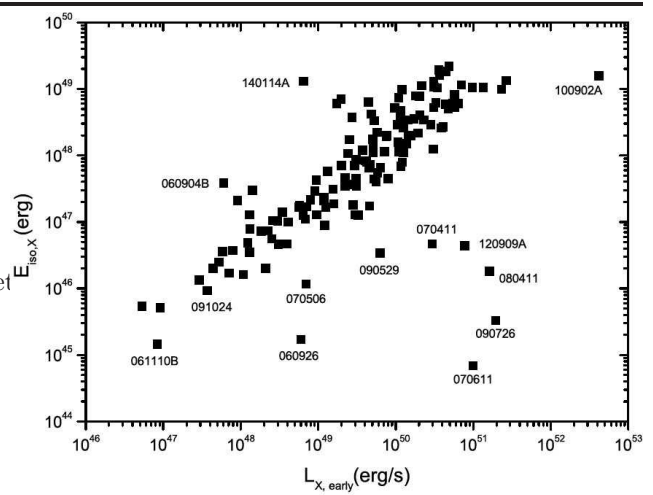}
\caption{\emph{Total isotropic X-ray afterglow energy $E_{iso,X}$
against the X-ray afterglow luminosity measured very early,
denoted as $L_{X,early}$. ($\Omega_k=0$, $\Omega_{\Lambda}=0.7$,
$\Omega_M=0.3$, $H_0= 70~km/s/Mpc$)}} \label{fig1azzam3}
\end{figure}

With this sample of 126 GRBs, we study the correlation between the
prompt gamma emission and the X-ray afterglow. In
Figure(\ref{figAazzam3}), we plot the isotropic X-ray afterglow
energy, denoted by $E_{iso,XA}$ against the gamma isotropic
emission, denoted by $E_{iso,\gamma}$. We confirm a correlation
between these quantities, which are obtained using the expressions
\eqref{eq.1b} and \eqref{eq.2}. The first correlation relation is
expressed analytically by the equation \eqref{eq.9}. We note that
we find exactly the same slope as was found by
\citep{margutti:13}. In a recent work, \cite{zaninoni:15} find the
following for long bursts: $m=0.68\pm0.06$ and $y_0 = 16\pm 2$,
corresponding to $q=-0.64\pm0.08$. In our work, we obtain the
following expression:
\begin{eqnarray}
\log{(\frac{E_{iso,XA}}{erg})}&=& (0.74\pm
0.05)~\log{(\frac{E_{iso,\gamma}}{~erg})}\nonumber\\
&+& (12.3\pm2.6).\label{eq.9}
\end{eqnarray}

\begin{figure}[h]
\centering
\includegraphics[angle=0, width=0.483\textwidth]{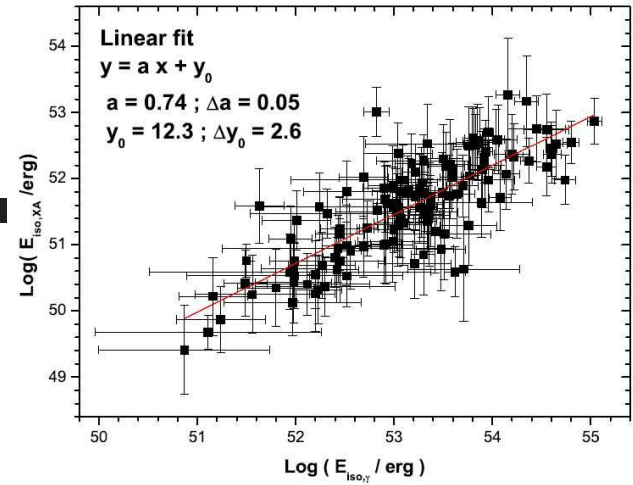}
\caption{\emph{Total X-ray afterglow isotropic energy,
$E_{iso,XA}$, against the prompt emission isotropic energy,
$E_{iso,\gamma}$, for our 126 GRBs. }} \label{figAazzam3}
\end{figure}

In Figure \ref{fig2bz} we present a histogram of our 126 GRBs as a
function of $\log{(E_{iso,XA} / E_{iso,\gamma})}$. We note that 67
GRBs (53 \% of the sample) have a ratio $r= (E_{iso,XA} /
E_{iso,\gamma}) <  3\%$, and 124 of them (98.4 \% of the sample)
have $r < 30 \%$. On average, we find $r = 0.03^{+0.07}_{-0.02}$.
Considering the error box, this result is in agreement with the
ratio of $10 \%$ obtained by \cite{willingale:07} and confirmed by
\cite{dainotti:15b}.

\begin{figure}[h]
\centering
\includegraphics[angle=0, width=0.483\textwidth]{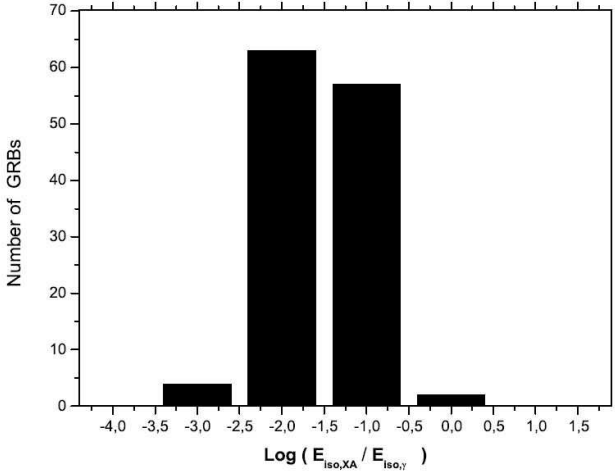}
\caption{\emph{Histogram of our 126 GRBs as a function of
$\log{(E_{iso,XA} / E_{iso,\Gamma})}$ with a logarithmic step
equal to 1. }} \label{fig2bz}
\end{figure}

Aiming to confirm a correlation relation between an observed
quantity and an intrinsic source quantity, we have studied the
correlation relations between $L_X(T_a)$, the X-ray afterglow
luminosity determined at the time of the break $T_a$ in the
temporal profile of the X-ray flux after the plateau, and the
break time $T_a$ itself. We thus needed to determine those breaks
in the X-ray flux time profiles, the latter being obtained from
the \textit{Swift} database \textit{Swift}/XRT\citep{evans:09}.

We should note that a correlation between $T_a$ and $L_X(T_a)$ was
found by \cite{dainotti:08} based on a sample of 32 GRBs detected
by \textit{Swift}. The discovery of this relation has been the
object of several updates based on newer sets of data
\citep{{dainotti:10},{dainotti:11},{dainotti:13},{dainotti:13b},{dainotti:15a}}.
These authors expressed $L_X$ as function of the break time
$T_a/(1+z)$ in the source frame (the logarithmic variable
$\log{(T_a/(1+z))}$. Using the maximum likelihood estimator,
\cite{dainotti:15b} get $m = -0.90^{+0.19}_{-0.17}$ and $ q =
51.14\pm0.58$ on a sample of 123 GRBs. In a recent work,
\cite{dainotti:16} add a third parameter, $L_{peak}$, and infer a
new correlation plane from a total sample of 176 \textit{Swift}
GRBs. In our work, we have preferred to keep $L_X$ as the
independent variable and study a correlation with $T_a$.

Following that, we study the different possible correlations
between the various intrinsic physical quantities obtained in the
source's reference frame ($L_{X}(T_a)$, $E_{iso,XA}$,
$E_{iso,\gamma}$) and the quantities observed by
\textit{Swift}/XRT, i.e. the break time $T_a$ and the X-ray flux
$FX(T_a)$. We also search for a correlation between observed flux
$FX(T_a)$ and intrinsic break time $T_a/(1+z)$.

This study is performed on a sample of 73 GRBs, after having kept
only those bursts with an X-ray flux that has a plateau followed
by a break and which can be fitted by the phenomenological model
given by \citep{willingale:07}.

In Table \ref{tabzh1zga} we list the 73 GRBs with their redshifts
and their characteristic quantities which we have calculated. The
temporal profiles of these bursts more or less resemble those
shown in Figure \ref{fig1azzam3b}, which presents the profile of
the most recent GRB in our sample, i.e. GRB150323A.

To find the break time $T_a$ and its uncertainty, we use the data
given in the official Swift/XRT
website\footnote{http://www.swift.ac.uk/xrt\_live\_cat/}, which
automatically treats the raw data; it classifies bursts into 5
types: (a) canonical; (b) one break, step first; (c) one break,
shallow first; (d) no breaks; (o) oddball. Out of the 126 GRBs, we
find 65 of type (a), 16 of type (b), 6 of type (c), 6 of type (d),
and 31 of type (o). We use only the ``canonical" ones (type a) and
those that have a break after the X-ray flux plateau. We also add
two bursts of type O, GRB100413A and GRB120729A, as their profiles
are very similar to the (a) type. We thus end up with a sample of
73 bursts, given in a Table \ref{figAazzam3}.

\begin{figure}[h]
\centering
\includegraphics[angle=0, width=0.483\textwidth]{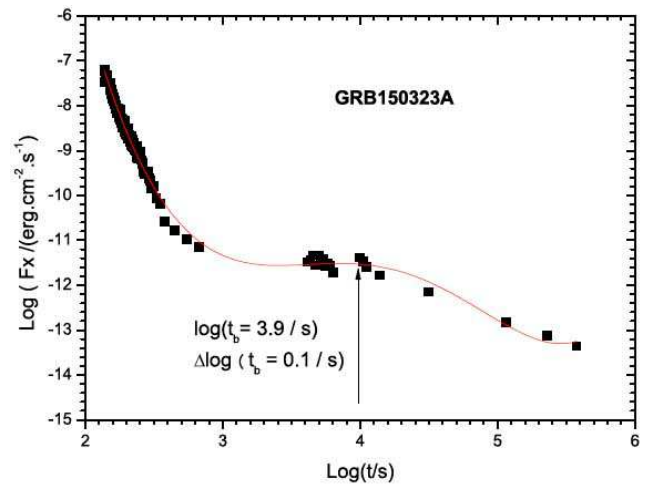}
\caption{\emph{Time profile of the X-ray afterglow flux of
GRB150323A.}} \label{fig1azzam3b}
\end{figure}

In Figure (\ref{fig2azzam3}) we present the relation between the
X-ray flux and the break time $T_a$ for our sample of 73 GRBs. We
confirm a correlation between the two quantities, and we express
that with Equation \eqref{eq.1}. This formula does not allow us to
constrain the cosmological parameters, because both quantities are
observed and independent of these parameters, but it does
encourage us to try to confirm a correlation between $T_a$ and the
luminosity $L_X$ at $T_a$.

We have thus sought such a correlation in a flat universe
($\Omega_k=0$, $\Omega_{\Lambda}=0.7$, $\Omega_M=0.3$, $H_0=
70~km/s/Mpc$). The correlation is shown graphically in Figure
\ref{fig4azzam3} and analytically by Equation \eqref{eq.3}. We
note that these relations have rather good precisions, judging by
the values of their slopes and intercepts  (see the uncertainties
on the power indices in Equations \eqref{eq.1} and \eqref{eq.3}).
\begin{eqnarray}
F_X(T_a)&=& 10^{-7.02\pm0.35}~T_a^{-1.01\pm0.08},\label{eq.1}\\
\frac{T_a}{1+z}&=& 10^{25.2\pm 1.8}
~~L_X(T_a)^{-0.46\pm0.04},\label{eq.3}
\end{eqnarray}
with $F_X$ in $(erg/cm^2/s)$, $ T_a$ in seconds, and $L_X$ in
$erg/s$.  z is the redshift and $T_a/(1+z)$ is the break time
measured in the source's rest frame.

We may compare with the recent work of \cite{hendrik:14}, who
finds a slope of $-1.07^{+0.20}_{-0.09}$ for $F_X$ as a function
of $T_a$,  which is quite close to our own result.

\begin{figure}[h]
\centering
\includegraphics[angle=0, width=0.483\textwidth]{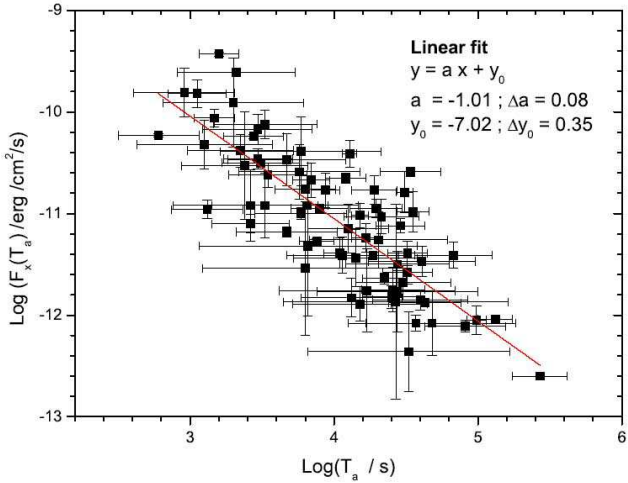}
\caption{\emph{X-ray flux calculated at the break time as a
function of the break time $T_a$ for 73 GRBs. ($\Omega_k=0$,
$\Omega_{\Lambda}=0.7$, $\Omega_M=0.3$, $H_0= 70~km/s/Mpc$)}}
\label{fig2azzam3}
\end{figure}

\begin{figure}[h]
\centering
\includegraphics[angle=0, width=0.47\textwidth]{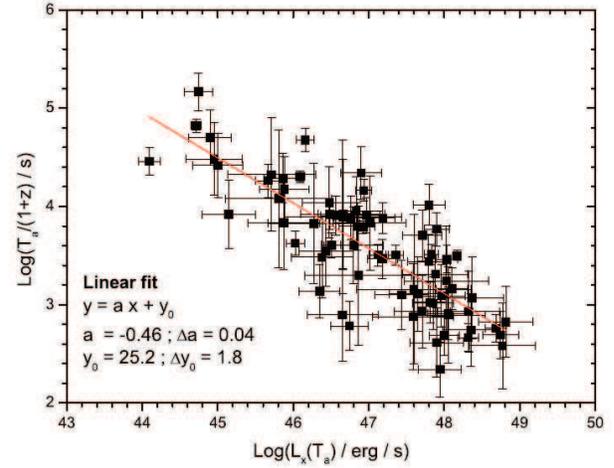}
\caption{\emph{Luminosity versus break time for 73 GRBs.
($\Omega_k=0$, $\Omega_{\Lambda}=0.7$, $\Omega_M=0.3$, $H_0=
70~km/s/Mpc$)}} \label{fig4azzam3}
\end{figure}

 For that same sample of 73 GRBs we have studied
the correlation between $L_X(T_a)$ and the isotropic energy of the
prompt gamma emission, $E_{iso,\gamma}$.  We present this relation
graphically in Figure(\ref{fig5azzam3}) and analytically by
Equation \eqref{eq.13}. This relation has an uncertainty of $16\%$
on the slope and about $100\%$ on the value of the intercept.
\begin{figure}[h]
\centering
\includegraphics[angle=0, width=0.483\textwidth]{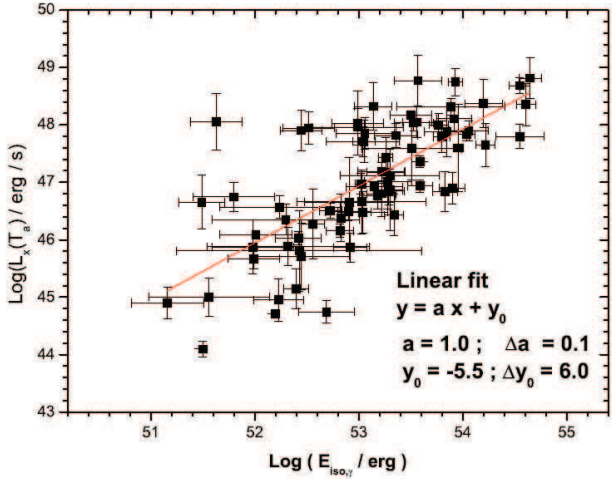}
\caption{\emph{X-ray afterglow luminosity, $L_X$, calculated at
the break time as a function of $E_{iso,\gamma}$ for our sample of
70 GRBs. ($\Omega_k=0$, $\Omega_{\Lambda}=0.7$, $\Omega_M=0.3$,
$H_0= 70~km/s/Mpc$)}} \label{fig5azzam3}
\end{figure}

\begin{equation}
L_X(T_a)= 10^{-5.5\pm6.0}~E_{iso,\gamma}^{1.0^\pm0.1}.\label{eq.13}\\
\end{equation}

We note that this correlation relation suffers from a very large
uncertainty on the value of the intercept, therefore it cannot be
 used to make any convincing inferences. By contrast, despite substantial scatter, the $L_X - T_a$
  plot gives an intercept with only ~ 10 \% uncertainty. From the correlations that
we have sought, we thus only retain the one between the break time
$T_a$ and the luminosity $L_X$ at that time, with the goal of
constraining cosmological parameters.

\section{Cosmological parameters derived from correlation relations}

We use the maximum likelihood method as described in
\citep{{dagostini:05}, {amati:08}, {dainotti:13}, {dainotti:16}}
to constrain the cosmological parameters within the standard
$\Lambda CDM$ model. We should note that in this work we try to
constrain the cosmological constants $\Omega_{\Lambda}$ and
$\Omega_M$ while taking a value $H_0 = 70~km/s/Mpc$ for the Hubble
constant.

\subsection{Usage of the Amati Relation}

We start by using the Amati relation, as presented in our previous
work \citep{zitouni:14}, in an effort to constrain the
cosmological constant $\Omega_M$. The Amati relation is given by
the following equation:
\begin{equation}
\frac{E_{p,i}}{keV}=K\times (\frac{E_{iso}}{10^{52} erg})^m,
\end{equation}
where $E_{p,i}$ is the energy of the burst corresponding to the
peak of the flux and measured in the source's frame, and $E_{iso}$
is the total energy emitted by the source in all space. In this
work, we use data for 27 bursts to infer the constants $K$ and $m$
\citep{zitouni:14}, and we assume a flat universe, such that
$\Omega_{M}+\Omega_{\Lambda}=1$. It thus suffices to constrain one
parameter to obtain the other.

Originally, the Amati relation was discovered for a flat universe,
characterized by $\Omega_k=0$, $\Omega_{\Lambda}=0.7$,
$\Omega_M=0.3$, $H_0= 70~km/s/Mpc$. These values  were chosen
based on the results obtained from SNe Ia data. In Table
\ref{tab1}, we present the values of the $m$ and $q$ fitting
parameters found in the framework of this model.

{\scriptsize {
\centering
\begin{table}[h]
\centering \caption{\emph{Values of the slope m and intercept q.
To compare with the original Amati relation, one must take $q =
\log{K}-52~m $. Ref(1): Present work, Ref(2):\cite{amati:03},
Ref(3): \cite{ghirlanda:04}. ($\Omega_k=0$,
$\Omega_{\Lambda}=0.7$, $\Omega_M=0.3$, $H_0= 70~km/s/Mpc$)}}
\label{tab1}
\begin{tabular}{|c|c|c|c|}
\hline
$m$ & $q$ & $\sigma_{int}$ & Ref.\\
\hline
$0.37\pm0.07$ & $-15\pm 3$ & $0.20\pm0.01$& (1)\\
$0.35\pm0.06 $ & $-16\pm3$ & ~~&(2) \\
$0.40\pm0.05$ &$ -18.8\pm2.7$ &~~&(3)\\
\hline
\end{tabular}
\end{table}
}}

We stress that our novel approach consists in performing a
``reverse job", namely to search for the cosmological parameters
that give the best fit, which is when the likelihood function
$-\ln{\mathcal{L}}$ is minimized. The fit is not given by specific
values but rather by surfaces or contours corresponding to the
same values of $-\ln{\mathcal{L}}$. We vary the cosmological
constant $\Omega_M$ numerically between 0 and 1, and determine the
values presented in Table \ref{tab2} for the Amati correlation
relation.

{\scriptsize {
\centering
\begin{table}[h]
\centering \caption{\emph{Best values obtained for the parameters
of the Amati relation using the likelihood method. Data used here
consist of 27 GRBs, taken from our previous work,
\citep{zitouni:14}. Values are obtained assuming a flat universe
($\Omega_k = 0$), with $\Omega_M$ varying between $0$ and $1$ and
$\Omega_{\Lambda }= 1- \Omega_M$.} } \label{tab2}
\begin{tabular}{|c|c|c|}
\hline
parameters & $min$ & $max$ \\
\hline
m & 0.32 & 0.37\\
q & -17 & -15 \\
$\sigma_{int}$ &0.19&0.21\\
-$\ln{\mathcal{L}}$&-25&-23.5\\
\hline
\end{tabular}
\end{table}
}}

In Figure \ref{fig7azzam3}, we plot the values of the function
$-ln{\mathcal{L}}$ in the ($m,\Omega_M$) plane for a constant
value of $\sigma_{int}=0.2$. Each contour is characterized by one
value of $-ln{\mathcal{L}}$. We note that the best value for
$-ln{\mathcal{L}}$ corresponds to  low values of $\Omega_M \leq$.

\begin{figure}[h]
\centering
\includegraphics[angle=0, width=0.48\textwidth]{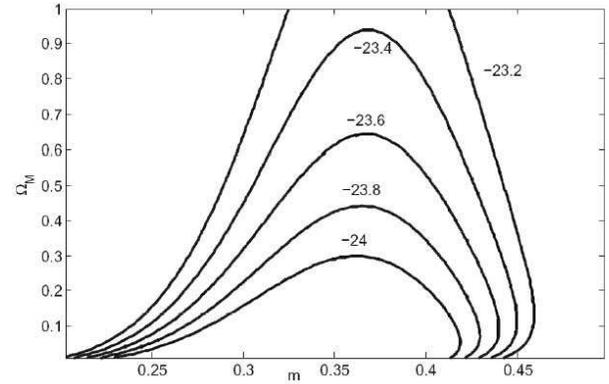}
\caption{\emph{The function $-\ln{\mathcal{L}}$ in the plane ($m$
, $\Omega_M$), with $\sigma_{int} = 0.20$ for the Amati relation
obtained from the data for 27 GRBs of \textit{Swift}/BAT
\citep{zitouni:14}. A contour represents the same value of
$-\ln{\mathcal{L}}$ for different pairs of ($m,\Omega_M$). We
assume a flat universe, i.e. $\Omega_k = 0.$} } \label{fig7azzam3}
\end{figure}

In Figure \ref{fig8azzam3b}, we plot the values of the function
$-ln{\mathcal{L}}$ in the ($\sigma_{int},\Omega_M$) plane for a
slope value of $m = 0.37$. Each contour is characterized by one
value of $-ln{\mathcal{L}}$. We note that the best value for
$-ln{\mathcal{L}}$ corresponds to low values of $\Omega_M \leq
0.3$.

\begin{figure}[h]
\centering
\includegraphics[angle=0, width=0.48\textwidth]{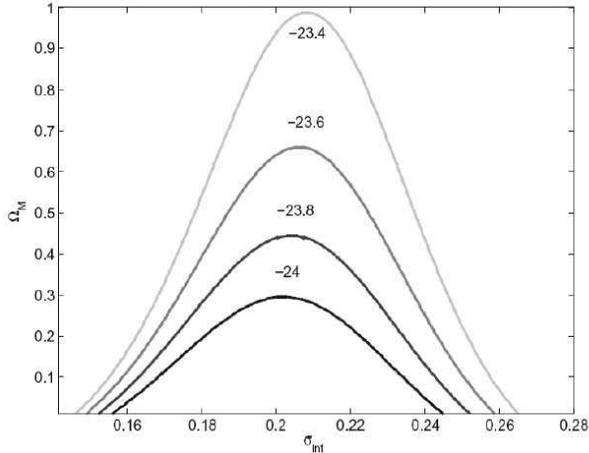}
\caption{\emph{The function $-\ln{\mathcal{L}}$ in the
($\sigma_{int}$, $\Omega_M$) plane, with a slope $m = 0.37$, for
the Amati relation obtained from the data for 27 GRBs in
\textit{Swift}/BAT \citep{zitouni:14}. A contour represents the
same value of $-\ln{\mathcal{L}}$ for different pairs of
($\sigma_{int}$, $\Omega_M$). We assume a flat universe, i.e.
$\Omega_k = 0.$}} \label{fig8azzam3b}
\end{figure}

In Figure \ref{fig7azzam3c}, we plot values of the function
$-ln{\mathcal{L}}$ in the plane ($m,\sigma_{int}$), assuming a
constant value of $\Omega_M = 0.3$. Each contour represents one
value of $-ln{\mathcal{L}}$. We note that the best fit is for
$-ln{\mathcal{L}} = -24$, which corresponds to a slope $m =
0.360\pm 0.005$ and $\sigma_{int} = 0.202 \pm 0.004$. The
intercept of the best fit is $-17\pm1$.

\begin{figure}[h]
\centering
\includegraphics[angle=0, width=0.48\textwidth]{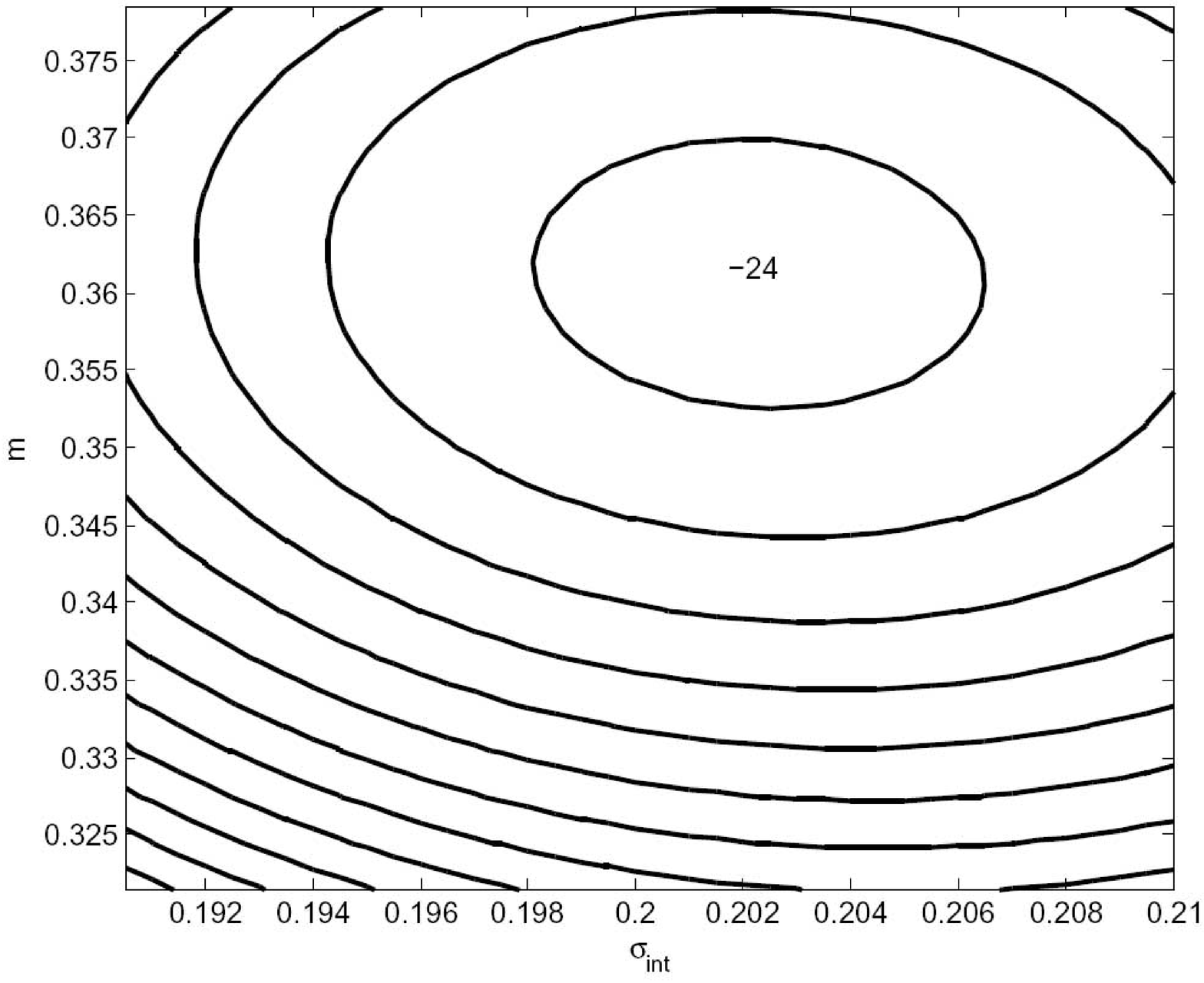}
\caption{\emph{The likelihood function in the ($m,\sigma_{int}$)
plane for $\Omega_{M} = 0.2975$, $\Omega_{\Lambda} = 0.70$ and
$\Omega_k = 0$ for the Amati relation obtained from the
\textit{Swift}/BAT data. The contours represent the same value of
the function $-\ln{\mathcal{L}}$ obtained for different pairs ($m$
,$\sigma_{int}$). For the innermost contour,
$-\ln{\mathcal{L}}=-24$, and for each next one, 0.02 is added.}}
\label{fig7azzam3c}
\end{figure}

Next, we attempt to constrain the cosmological parameters
$\Omega_{\Lambda}$ and $\Omega_M$, using the Amati relation as
studied in our previous work \citep{zitouni:14}, but in any
universe (cosmological topology), that is assuming a curvature
constant $\Omega_k = 1 -\Omega_M-\Omega_{\Lambda}$. We first
determine the best values of $L = -\ln{\mathcal{L}}$, which
correspond to the parameters $(m,q)$ of a straight line. In this
case we vary $\Omega_M$ between 0 and 1.2 and $\Omega_{\Lambda}$
between 0 and 1, independently. Our results are shown graphically
in Figure \ref{fig7azzam3cc} and (with more detail) in tabular
form in Table \ref{tab3}. The best values of the function
$-\ln{\mathcal{L}}$ correspond to small values of $\Omega_M$ and
large values of $\Omega_{\Lambda}$. In other words, the
statistical method that is used tends to favor a universe
dominated by dark energy if the Amati relation is correct, and not
simply due to a selection effect \citep{nakar:05}. For example,
for $\Omega_M = 0.0175 ; \Omega_{\Lambda}=0.975$ we get $m =
0.3275\pm 0.0025$ and $\sigma_{int} = 0.193\pm 0.0015$ with rather
high precision.

\begin{figure}[h]
\centering
\includegraphics[angle=0, width=0.48\textwidth]{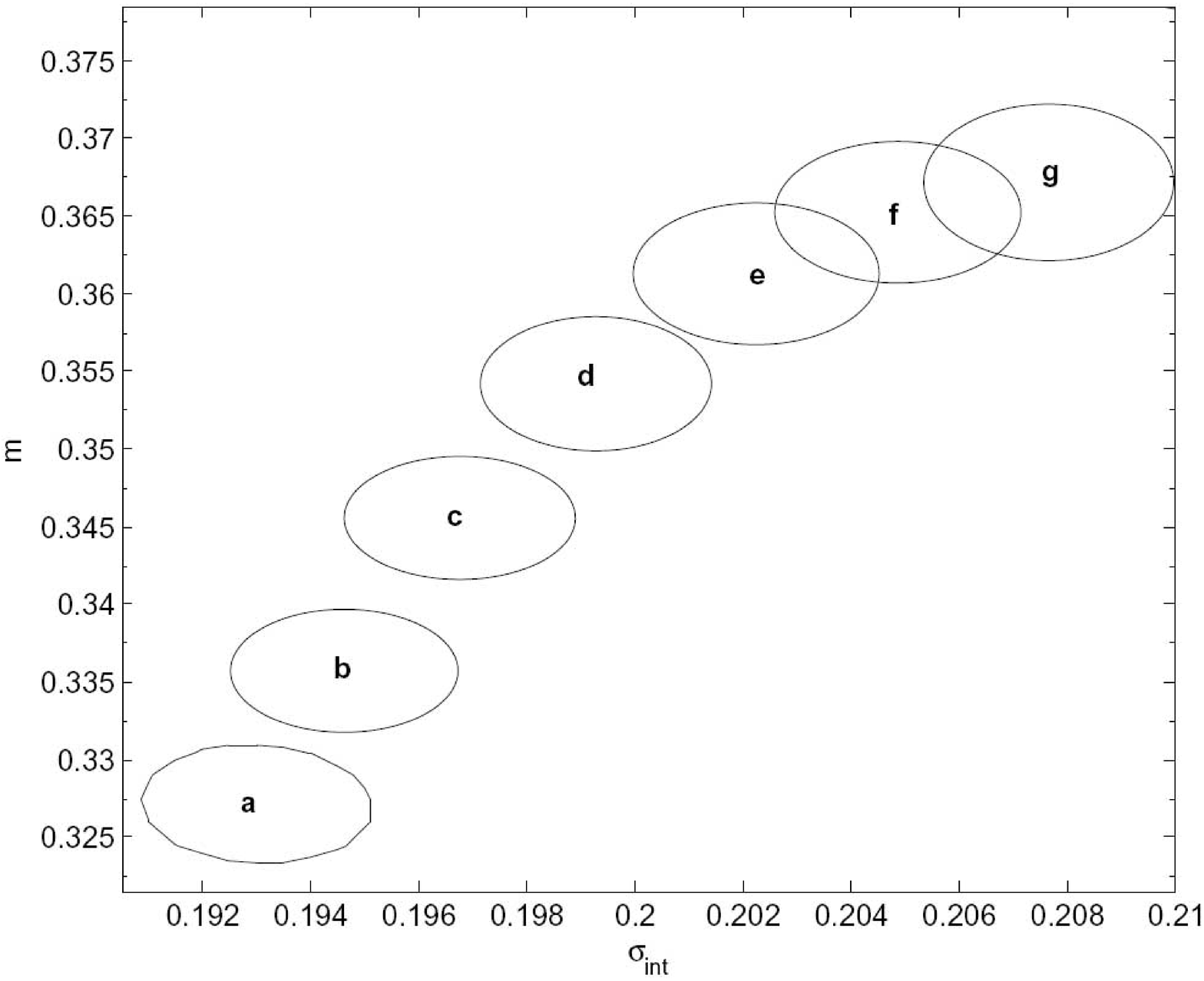}
\caption{\emph{The likelihood fonction in the ($m, \sigma_{int}$)
plane for different values of $\Omega_{M}$ and $\Omega_{\Lambda}$
using the Amati relation with \textit{Swift}/BAT data for 27 GRBs
\cite{zitouni:14}. The contours represent the values of the
$-\ln{\mathcal{L}}$ function. For more information, please refer
to Table (\ref{tab3}). }} \label{fig7azzam3cc}
\end{figure}

{\scriptsize {
\centering
\begin{table}[t]
\centering
\caption{Values of the slope $m$ and $\sigma_{int}$ for various
values of $\Omega_{M}$ and $\Omega_{\Lambda}$. a, b and f represent the contours of Figure (\ref{fig7azzam3cc}) }
\label{tab3}
\begin{tabular}{|c|c|c|c|}
\hline
&$\Omega_M$&$\Omega_{\Lambda}$&$-\ln{\mathcal{L}}$\\
\hline
a&0.0175& 0.975&-25.03\\
b&0.0525&0.975&-24.83\\
c&0.0875&0.925&-24.58\\
d&0.1575&0.80&-24.32\\
e&0.2975&0.70&-24.00\\
f&0.50&0.50&-23.74\\
g&0.10&0.997&-23.47\\
\hline
\end{tabular}
\end{table}
}}

In Figure \ref{fig7azzam4cc} we show the best values of the
function $-\ln{\mathcal{L}}$ plotted in an $\Omega_M$,
$\Omega_{\Lambda}$ plane diagram. On the same figure we present
the contours of the values obtained by our methods for determining
the cosmological parameters using the SNe Ia data. We note that it
is difficult to constrain the cosmological constants using GRB
correlation relations without making use of supernovae data.
\citep{{riess:04},{xu:05}}. The methods agree for values centered
around $\Omega_M = 0.3$ and $\Omega_{\Lambda} = 0.7$. We wish to
stress  the fact that we did not use the Amati relation as found
for the specific values of $\Omega_M = 0.3$ and $\Omega_{\Lambda}=
0.7$ in order to constrain these same parameters; that would be
falling into the circularity trap\citep{{ghirlanda:04b},{dai:04}}.

By inverting the Amati relation and taking $E_{p,i}$ as an
independent variable and $E_{iso}$ as a variable which depends on
the cosmological parameters, we obtain the results shown in Figure
\ref{fig20}. This second Amati relation is expressed by
$\log{E_{iso}} = m~\log{E_{p,i}} + q$. In this case, the
corresponding values of $m$, $q$, $\sigma_{int}$, and
$-ln{\mathcal{L}}$ are given in Table \ref{tab5}. We note that the
values of $\sigma_{int}$ and $-\ln{\mathcal{L}}$ are larger than
those obtained with the first Amati relation. On the other hand,
we note that with the inverse Amati relation, the likelihood
methods tends to prefer cosmological parameters that converge
toward $\Omega_{M} = 0.28$ et  $\Omega_{\Lambda} = 0.725$.

{\scriptsize { \centering
\begin{table}[h]
\centering \caption{\emph{Best values obtained for the parameters
of the second Amati relation using the likelihood method for
various values of $\Omega_{M}$ and $\Omega_{\Lambda}$. Data used
here consist of 27 GRBs, taken from our previous work,
\citep{zitouni:14}.}}  \label{tab5}
\begin{tabular}{|c|c|c|c|}
\hline
$m$ & $q$ & $\sigma_{int}$ & $-ln{\mathcal{L}}$ \\
\hline
1.15 &49.72& 0.4216 & -3.79\\
1.20 &49.71& 0.4240 & -3.90\\
1.25 &49.65& 0.4264 & -3.94\\
1.30 &49.75& 0.4328 & -3.87\\
1.45 &49.56& 0.4512 & -3.30\\
 \hline
\end{tabular}
\end{table}
}}

Our approach was rather to find the best values of the
cosmological parameters that correspond to the minimum value(s) of
($-\ln{\mathcal{L}}$) and to then infer the correlation constants
$m$ and $q$. This method, however, is very sensitive to the
dispersion of the data. It allows one to converge on rather
precise values if the data have been obtained with high precision.
This procedure also allows one to verify a correlation relation by
comparing with the results obtained through other methods and data
(WMAP: the Wilkinson Microwave Anisotropy Probe, SN Ia).

\begin{figure}[h]
\centering
\includegraphics[angle=0, width=0.48\textwidth]{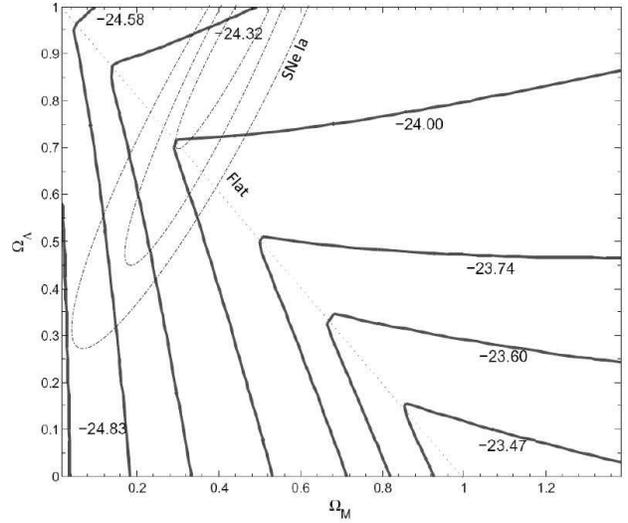}
\caption{\emph{The contours of the function $-ln(\mathcal{L})$ in
the $\Omega_{M}, \Omega_{\Lambda}$ plane, using the Amati relation
obtained from the \textit{Swift}/BAT data. The contours represent
the same values of the function $-\ln{\mathcal{L}}$ corresponding
to $m$ and $\sigma_{int}$ given in Table \ref{tab4}. The
dotted-line contours are the results obtained using the methods
based on SNe Ia.}} \label{fig7azzam4cc}
\end{figure}

\begin{figure}[h]
\centering
\includegraphics[angle=0, width=0.48\textwidth]{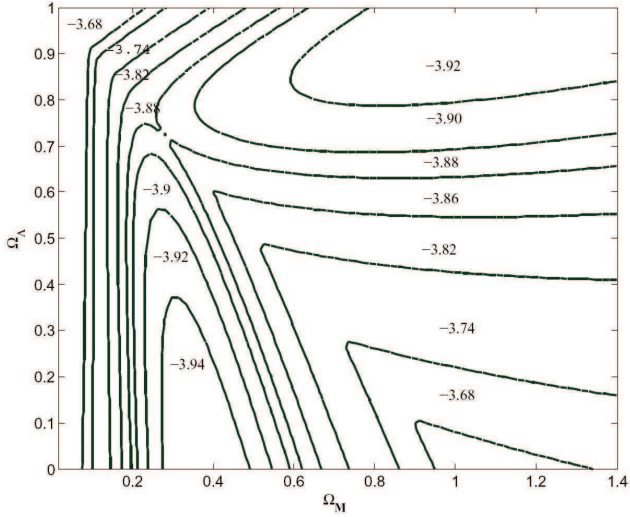}
\caption{\emph{The contours of the function $-ln(\mathcal{L})$ in
the $\Omega_{M}, \Omega_{\Lambda}$ plane, using the inverted Amati
relation obtained from the \textit{Swift}/BAT data. The contours
represent the same values of the function $-\ln{\mathcal{L}}$
corresponding to $m$ and $\sigma_{int}$ given in Table \ref{tab5}.
The meeting point of the contours correspond  $\Omega_{M}$= 0.28
and  $\Omega_{\Lambda}$ = 0.725}} \label{fig20}
\end{figure}

\subsection{Usage of the Dainotti relation}

We have also attempted to use the correlation that was obtained
above between the break time $T_{a}$ seen in the X-ray afterglow's
time evolution and the luminosity $L_X$ at that instant. For that
we used the 73 bursts that we had selected. We chose this
correlation  because it relates an observed quantity to  one which
is calculated in terms of the cosmological parameters. That
relation is expressed by Equation \eqref{eq.3} and applies to a
flat universe with ($\Omega_k=0$, $\Omega_{\Lambda}=0.7$,
$\Omega_M=0.3$, $H_0= 70~km/s/Mpc$). In what follows we study that
relation for a more general case.

Before applying our method, let us explain what we would like to
accomplish, assuming the ideal case in which our data do not
suffer from any dispersion. In Figure \ref{fig18bazzam3}, we
present the case where $\Omega_{M} = 0.3$. We note that our method
does converge toward ($\Omega_{M} = 0.3$, $\Omega_{\Lambda} =
0.7)$. This is a way to check the validity of our method and the
impact of the dispersion of data on our results.

\begin{figure}[h]
\centering
\includegraphics[angle=0, width=0.482\textwidth]{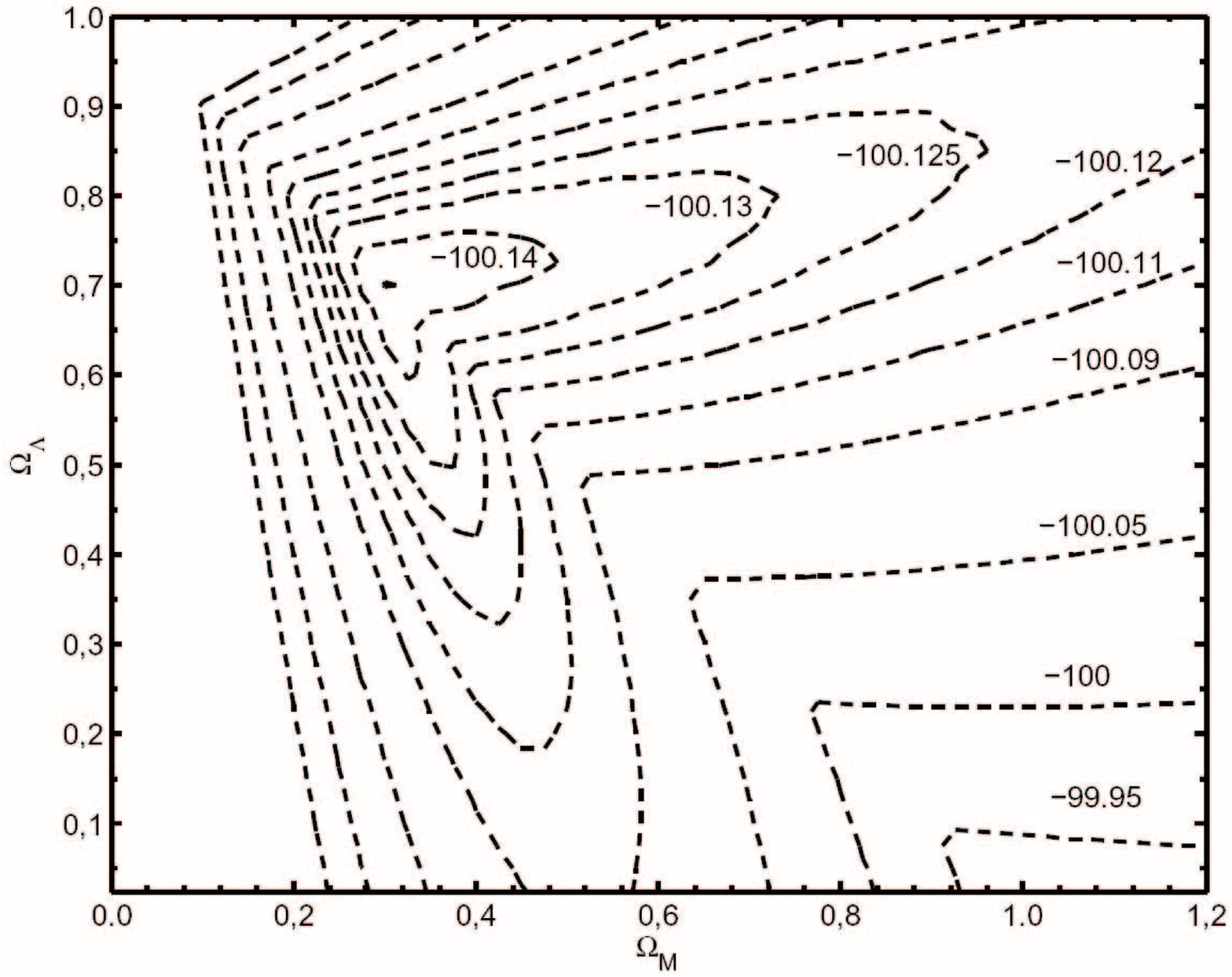}
\caption{\emph{The function $-\ln{\mathcal{L}}$ in the
($\Omega_{M}, \Omega_{\Lambda}$) plane for Dainotti correlation
relation between the X-ray luminosity X at $T_a$ as a function of
the break time after the temporal plateau. In the ideal case, we
have used a straight line ($\log{\frac{T_a}{1+z}}= -0.46
\log{F_x(T_a)} + 25.2$). }} \label{fig18bazzam3}
\end{figure}

In Figure \ref{fig17azzam3kk} we use the Dainotti correlation
relation $L_X(T_{a})-T_{a}$ to constrain $(\Omega_{M},
\Omega_{\Lambda})$ in any type of universe. We present the results
as contours of specific values of $-\ln{\mathcal{L}}$. We note
that the best values of this function, that is the minima of the
function, are obtained for $\Omega_{\Lambda} \rightarrow 0$ and
$\Omega_{M} \rightarrow 1$. In other words, Dainotti correlation
relation works best in a universe dominated by matter. This result
is opposite to what was obtained with the Amati relation. On the
other hand, if we include the results obtained using  supernovae,
we obtain the same earlier results, namely values closer to
$(\Omega_{M} =0.3, \Omega_{\Lambda} = 0.7)$ for a flat universe.

\begin{figure}[h]
\centering
\includegraphics[angle=0, width=0.48\textwidth]{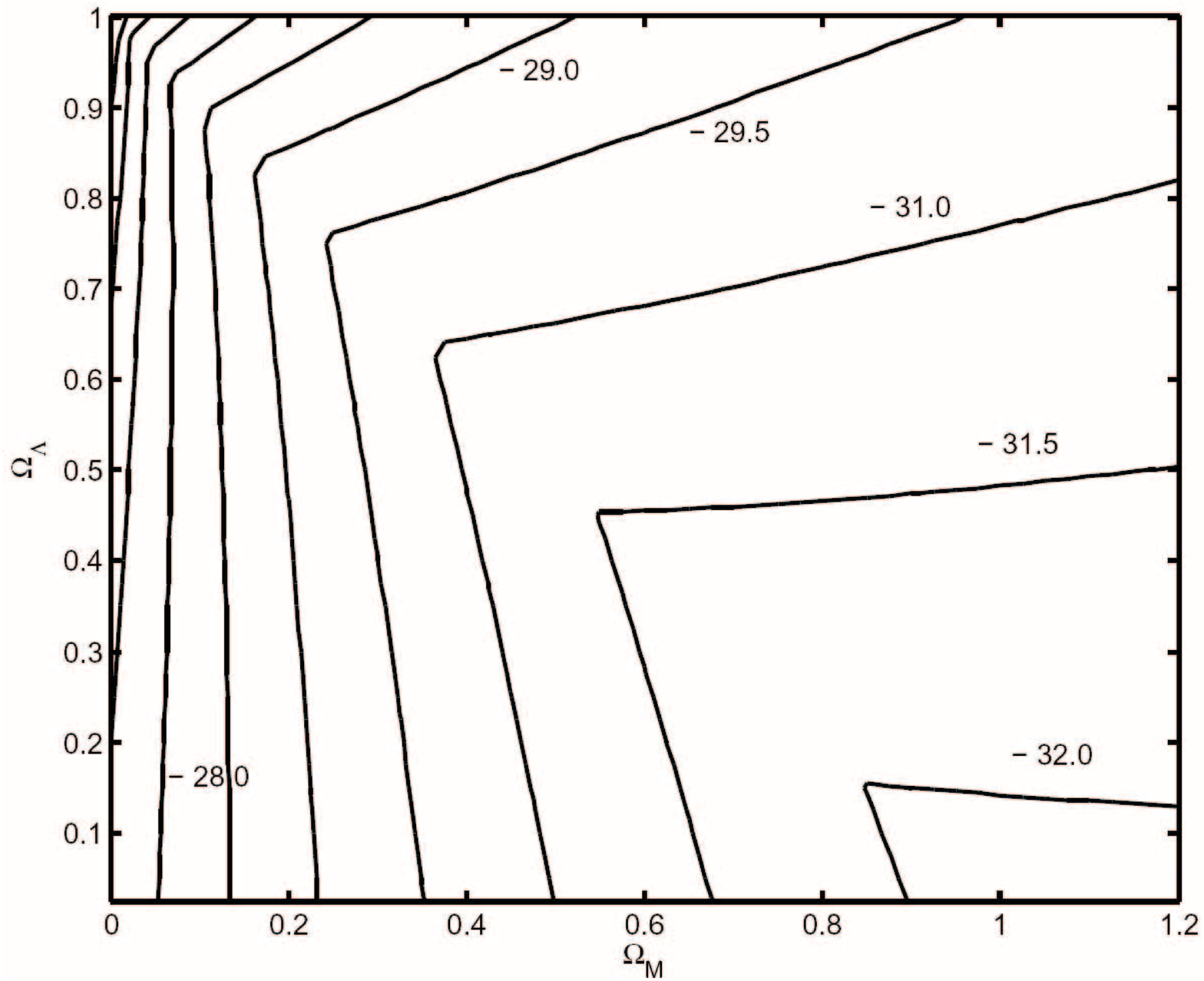}
\caption{\emph{The function $-\ln{\mathcal{L}}$ in the
($\Omega_{M}, \Omega_{\Lambda}$) plane using the Dainotti
correlation relation between $L_X(T_a)$ and $T_a$}}
\label{fig17azzam3kk}
\end{figure}

In Figure \ref{fig7azzam3dd} we show the contours for the best
values of the function $-\ln{\mathcal{L}}$ in the
$(m,\sigma_{int})$ plane for different values of the pair
$(\Omega_{M}, \Omega_{\Lambda})$. Information for each contour is
given in Table \ref{tab4}. For example, for contour B,
corresponding to $(\Omega_{M} =0.3, \Omega_{\Lambda} = 0.7)$, we
get $m = -0.462\pm 0.014$ and $ \sigma_{int} = 0.288\pm0.012$.

{\scriptsize {
\centering
\begin{table}[h]
\centering
\caption{Values for the slope $m$ and $\sigma_{int}$ for different values of $\Omega_{M}$ and $\Omega_{\Lambda}$. A and E represent the contours in Figure \ref{fig7azzam3dd}.}
\label{tab4}
\begin{tabular}{|c|c|c|c|}
\hline
&$\Omega_M$&$\Omega_{\Lambda}$&$-\ln{\mathcal{L}}$\\
\hline
A&0.90& 0.01&-32.02\\
B&0.50&0.50&-31.34\\
C&0.30&0.70&-30.72\\
D&0.10&0.90&-29.40\\
E&0.00&0.025&-28.55\\
\hline
\end{tabular}
\end{table}
}}

\begin{figure}[h]
\centering
\includegraphics[angle=0, width=0.48\textwidth]{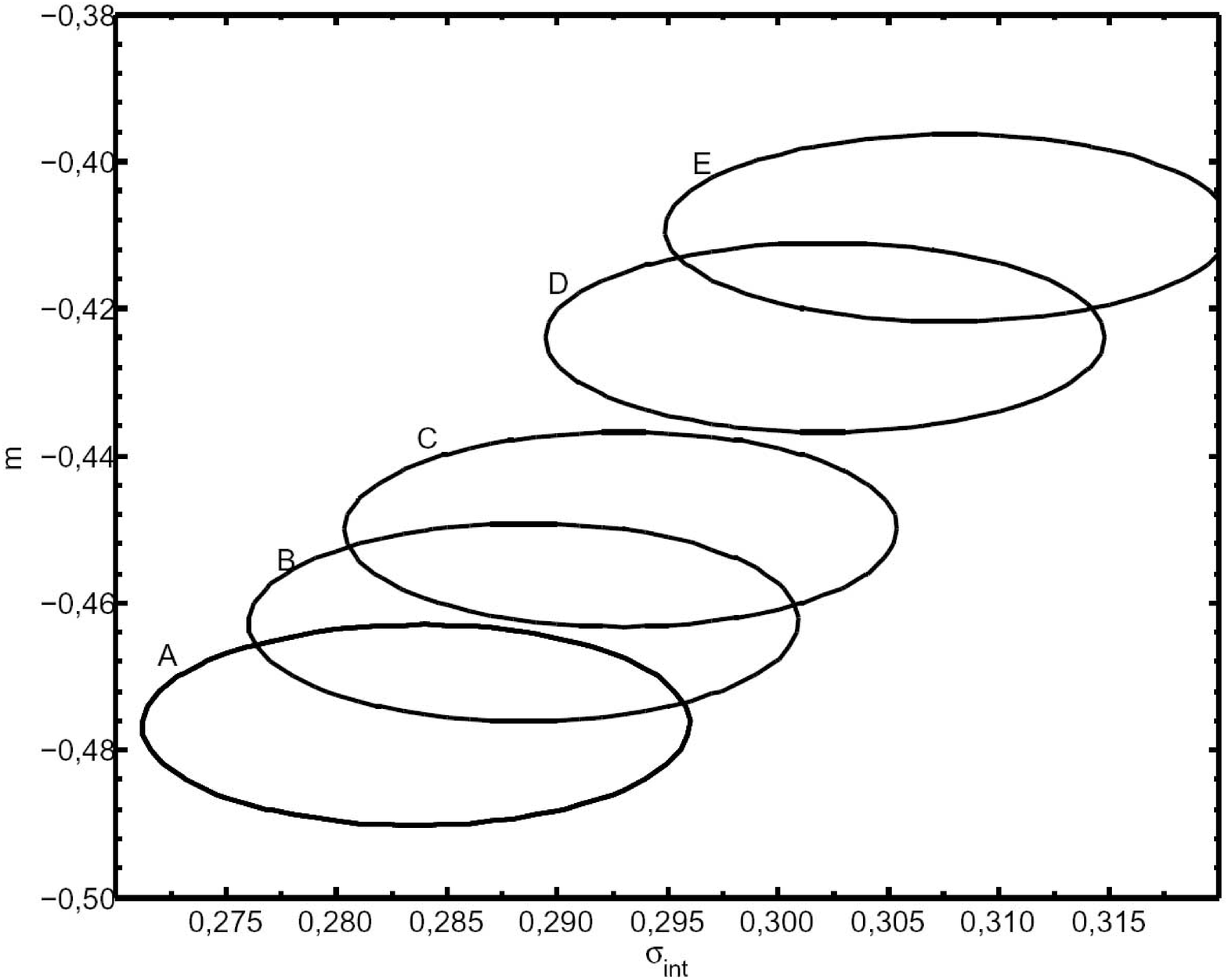}
\caption{\emph{The likelihood function in the ($m$ ,
$\sigma_{int}$) plane for different values of $\Omega_{M}$ and
$\Omega_{\Lambda}$ using the Dainotti correlation  between
$L_X(T_a)$ and $T_{a}$ we obtained from the \textit{Swift}/BAT
data for 70 GRBs. The contours represent the values of the
function $-\ln{\mathcal{L}}$. Further  information on A and E,
please refer to Table \ref{tab4}. }} \label{fig7azzam3dd}
\end{figure}

In Figure \ref{fig19azzam3} we show the obtained values of
$(m,\sigma_{int})$ starting from a flat space characterized by
$\Omega_M =0.3$. We here note the ability of the likelihood method
to converge to the best values of $m$ and $\sigma_{int}$.

\begin{figure}[h]
\centering
\includegraphics[angle=0, width=0.48\textwidth]{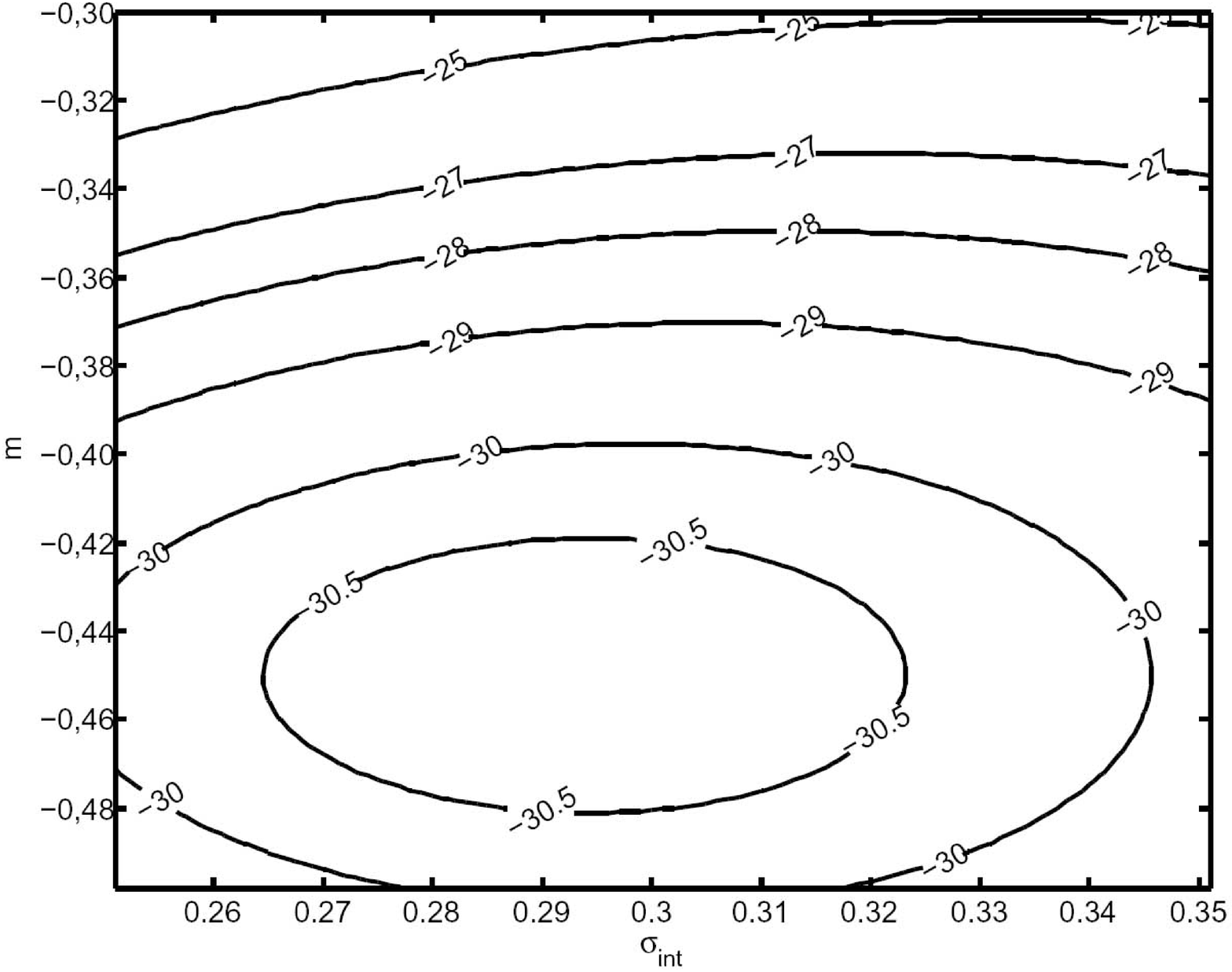}
\caption{\emph{The function $-\ln{\mathcal{L}}$ in the
($\sigma_{int}, m$) plane for Dainotti correlation  between the
X-ray luminosity and $T_a$ as a function of the break time after
the temporal plateau. $\Omega_M = 0.3$.}} \label{fig19azzam3}
\end{figure}

 When we express the Dainotti relation by taking $Ta/(1+z)$ as an
independent variable and $L_X$ as a variable that depends on
cosmological parameters, we obtain the results shown in Figure
\ref{fig21}. The relation that is represented there is
$\log{L_{X}} = m~\log{(T_{a}/(1+z))} + q$, and we refer to it as
the `second Dainotti relation' to distinguish it from the first
one referred to earlier. In this case, the corresponding values of
$m$, $q$, $\sigma_{int}$, and $-ln{\mathcal{L}}$ are given in
Table \ref{tab6}. We note that the values of $\sigma_{int}$ and
$-ln{\mathcal{L}}$ are larger than those obtained with the first
Dainotti relation while showing the same general trend.

{\scriptsize { \centering
\begin{table}[h]
\centering \caption{\emph{Best values obtained for the parameters
of the second Dainotti relation using the likelihood method for
various values of $\Omega_{M}$ and $\Omega_{\Lambda}$.}}
\label{tab6}
\begin{tabular}{|c|c|c|c|}
\hline
$m$ & $q$ & $\sigma_{int}$ & $-ln{\mathcal{L}}$ \\
\hline
-1.29 &51.38& 0.505 & 5.43\\
-1.30 &51.45& 0.510 & 6.05\\
-1.31 &51.53& 0.520 & 7.07\\
-1.33 &51.71& 0.535 & 9.05\\
-1.35 &51.89& 0.555 & 11.05\\
-1.38 &52.14& 0.590 & 14.09\\
-1.41 &52.37& 0.615 & 17.21\\
 \hline
\end{tabular}
\end{table}
}}

\begin{figure}[h]
\centering
\includegraphics[angle=0, width=0.48\textwidth]{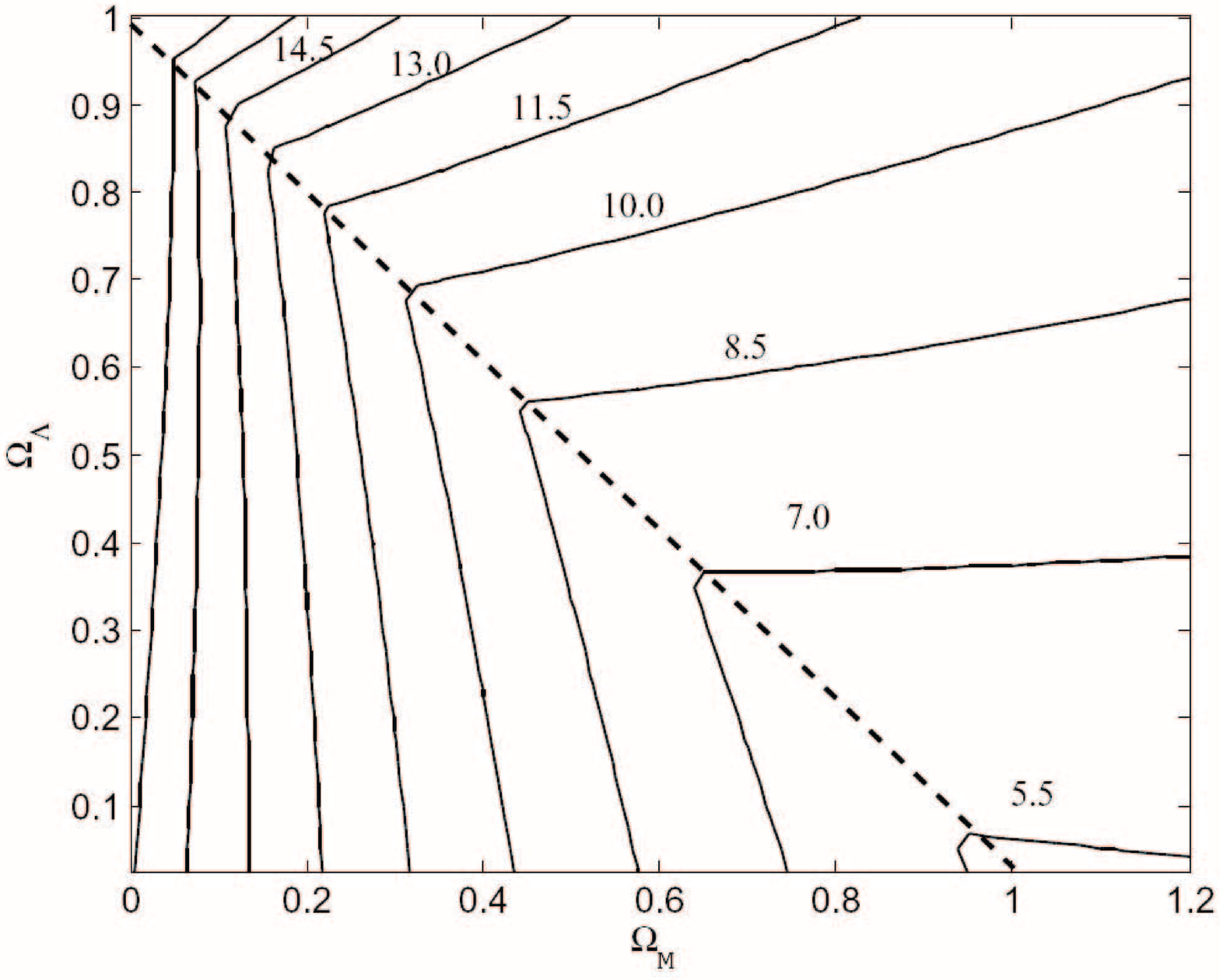}
\caption{\emph{The contours of the function $-ln(\mathcal{L})$ in
the $\Omega_{M}, \Omega_{\Lambda}$ plane, using the second
Dainotti relation obtained from the \textit{Swift}/BAT data. The
contours represent the same values of the function
$-\ln{\mathcal{L}}$ corresponding to $m$ and $\sigma_{int}$ given
in Table \ref{tab6}.}} \label{fig21}
\end{figure}

\section{Discussion}

This study was conducted to try to determine the cosmological
parameters $\Omega_{\Lambda}$ and $\Omega_M$ by using two
correlation relations: the Amati relation between $E_p$ and
$E_{iso}$ and a Dainotti relation between $T_a$ and $L_X(T_a)$,
which we presented in the previous sections. The Amati relation
has been widely used to this aim
\citep{{dai:04},{amati:08},{kodama:08},{wei:13},{wang:16}}.
However, this relation was inferred by assuming a flat universe
with $\Omega_M = 0.3$, while some of the above-mentioned works use
it as is and at the same time try to determine the cosmological
parameters, which raises the issue of the circularity problem.
Other works use a calibration based on the results from SN Ia for
redshifts $z < 1.414$ \citep{{wang:15},{wang:16}} , which we view
as a good approach (associating the results from SNe Ia with data
from GRBs).

In the present work, we have tried to constrain the cosmological
parameters using the two Amati relations relation but without
setting a priori values of the slope and intercept parameters m
and q. The best values are those that minimize the function $L =
-\ln{\mathcal{L}}$. This requirement leads to a set of values for
$\Omega_M$ et $\Omega_{\Lambda}$ that produce contours in the
plane of those cosmological variables/parameters. We find that the
first Amati relation tends to favor values of
$\Omega_{M}\rightarrow 0$ and $\Omega_{\Lambda}\rightarrow 1$. To
relate our result with that obtained from SNe Ia, we have
graphically superposed the two. On the other hand, the second
Amati relation favors values of the cosmological parameters that
tend to converge toward $\Omega_{M} = 0.28$ and $\Omega_{\Lambda}
= 0.725$ in the best cases.

 The second relation, Dainotti relation,  which we confirmed from \textit{Swift}/XRT data is
between the breaking time $T_a$ in the time profile of the X-ray
flux and the luminosity $L_X(T_a)$. However, it is characterized
by large dispersions of the data points around the interpolation
(straight) line. For the two versions of the Dainotti relation,
our statistical analysis tends to favor values of
$\Omega_{M}\rightarrow 1$ and $\Omega_{\Lambda}\rightarrow 0$. By
numerically reducing the dispersion of the data, the results are
greatly improved and converge to a set of values of the
cosmological parameters close to what is obtained by other methods

By numerically reducing the dispersion of the data, the results
are greatly improved and converge to a set of values of the
cosmological parameters close to what is obtained by other
methods; the results are also close to those presented in Figure
\ref{fig18bazzam3}, which seems to confirm the need for ``clean"
data with minimal data dispersion around the straight line. We
note that the Dainotti correlation has been used by several
authors in an effort to constrain the cosmological parameters
 \citep{{cardone:09}, {cardone:10}, {dainotti:13b}, {petrosian:15}}.

\section{Conclusion}
Gamma-ray bursts hold great potential as cosmological probes. This
fact, however, has not yet been fully utilized, mainly because
GRBs are not standard candles and partly because it has been
difficult to construct plots between GRB characteristics that do
not show too much scatter. The discovery and calibration of
several luminosity and energy correlations has ushered in a new
period of investigation in which GRBs are finally beginning to
prove their worth as cosmological probes. In this paper, we tried
to put limits on the values of  $q$  and  $m$ by utilizing the
well-known Amati relation and one that was obtained by
\cite{dainotti:08} from GRB data, namely a correlation between the
X-ray burst luminosity $L_X$ and the break time $T_a$ in the X-ray
flux's time profile, a correlation that we confirmed with our
sample. The latter relation suffers from wide scatter, but we were
able to narrow this scatter using numerical techniques. This
enabled us to get reasonable values for $\Omega_{\Lambda}$   and
$\Omega_M$ that are consistent with those obtained via other
methods. A few general conclusions may be drawn from this work:
\begin{enumerate}
\item  despite the wealth of GRB data that we now have (from \textit{Swift},
Fermi, and others), the data that is plotted in the "standard"
ways (luminosity vs. time, luminosity vs energy in various bands,
etc.) still shows much scatter, at least for cosmological research
purposes. Either we need more data in different energy bands or we
are missing some insights as to how to relate various quantities.
\item  The correlation that we have confirmed(between $L_X$ and $T_a$),
while far from perfect, shows that interesting perspectives can
still be obtained by looking at the data from different angles.
\item  Diversifying analysis approaches (maximum likelihood,
chi-square minimization, iterative convergence, etc.) can yield
interesting results that one may compare and contrast to reach the
most robust conclusions.
\item  For cosmological studies, while
GRBs may certainly represent an important new angle from which to
approach the determination of various parameters, combining
quantities and results from different methods (SN Ia supernovae,
Cosmic Microwave Background, Gamma Ray Bursts) and ensuring
consistency across the board appears to be not only the best
general approach but perhaps an absolutely necessary one.
\end{enumerate}

In the future, we hope to pursue this new, promising avenue along
the lines of the above general conclusions, in the aim of placing
more stringent limits on the values of cosmological parameters,
particularly by using larger data sets and GRB characteristics
(energies and fluxes from various intervals and bands), which may
aid in reducing the scatter in the correlation relations and thus
in obtaining more precise results.

\begin{acknowledgments}

 The authors gratefully acknowledge the use of the online \textit{Swift/BAT} table compiled by
 Taka Sakamoto and  Scott D. Barthelmy.  We thank the referee for very useful comments,
 which led to significant improvements of the paper.
\end{acknowledgments}
\bibliographystyle{spr-mp-nameyear-cnd}
\bibliography{zitouni_arxiv}
\appendix

{\scriptsize {
 \centering
\begin{longtable}{|c|c|c|c|c|c|c|c|c|}
\caption{Our sample consists of 73 GRBs, of which 65 are of type
(a) canonical; 6 are of type (c) one break, shallow first; 2 are
of type (o): oddball. $\Gamma$ is the X-ray spectral index. The
other quantities are defined in the text.}
\label{tabzh1zga}\\
  \hline
   GRB & z  &$\Gamma$&Log($\frac{E_{iso}}{\mathrm{erg}}$) & Log($\frac{E_{iso,XA}}{\mathrm{erg}}$) &
  Log($\frac{T_{a}}{\mathrm{s}})$&Log($\frac{FX(T_{a})}{\mathrm{erg/cm^{2}/s}}$)&Log($\frac{LX(T_{a})}{\mathrm{erg/s}})$\\
  \hline
\endfirsthead

\multicolumn{8}{c}%
{{\bfseries \tablename\ \thetable{} -- continued from previous
page}} \\ \hline
  GRB & z & $\Gamma$&Log($\frac{E_{iso}}{\mathrm{erg}}$) & Log($\frac{E_{iso,XA}}{\mathrm{erg}}$) &
  Log($\frac{T_{a}}{\mathrm{s}})$&Log($\frac{FX(T_{a})}{\mathrm{erg/cm^{2}/s}}$)&Log($\frac{LX(T_{a})}{\mathrm{erg/s}})$\\
  \hline
\endhead

\hline \multicolumn{8}{|c|}{{Continued on next page}} \\ \hline
\endfoot

\hline \hline
\endlastfoot
150323A $^a$    &   0.593   &   2.06    $\pm$   0.20    &   52.40   $\pm$   0.12    &   50.81   $\pm$   0.45    &   4.12    $\pm$   0.35    &   -11.83  $\pm$   0.19    &   45.15   $\pm$   0.35    \\
150314A $^c$    &   1.758   &   1.85    $\pm$   0.09    &   54.55   $\pm$   0.10    &   52.18   $\pm$   0.44    &   3.20    $\pm$   0.14    &   -9.43   $\pm$   0.02    &   48.69   $\pm$   0.14    \\
141121A $^a$    &   1.470   &   1.93    $\pm$   0.12    &   53.15   $\pm$   0.27    &   51.82   $\pm$   0.48    &   4.18    $\pm$   0.06    &   -11.02  $\pm$   0.12    &   46.92   $\pm$   0.06    \\
140907A $^a$    &   1.210   &   2.01    $\pm$   0.11    &   52.92   $\pm$   0.16    &   51.01   $\pm$   0.60    &   4.63    $\pm$   0.25    &   -11.88  $\pm$   0.02    &   45.86   $\pm$   0.25    \\
140703A $^a$    &   3.140   &   1.83    $\pm$   0.08    &   53.53   $\pm$   0.19    &   52.19   $\pm$   0.51    &   4.08    $\pm$   0.14    &   -10.65  $\pm$   0.05    &   48.03   $\pm$   0.14    \\
140419A $^a$    &   3.956   &   1.87    $\pm$   0.05    &   54.65   $\pm$   0.11    &   52.52   $\pm$   0.52    &   3.52    $\pm$   0.36    &   -10.12  $\pm$   0.14    &   48.81   $\pm$   0.36    \\
140304A $^a$    &   5.283   &   2.02    $\pm$   0.11    &   53.57   $\pm$   0.23    &   52.22   $\pm$   0.67    &   3.38    $\pm$   0.44    &   -10.53  $\pm$   0.53    &   48.77   $\pm$   0.44    \\
131103A $^a$    &   0.599   &   2.19    $\pm$   0.15    &   51.49   $\pm$   0.22    &   50.42   $\pm$   0.50    &   3.10    $\pm$   0.47    &   -10.32  $\pm$   0.24    &   46.65   $\pm$   0.47    \\
131030A $^a$    &   1.293   &   2.10    $\pm$   0.10    &   54.22   $\pm$   0.09    &   52.33   $\pm$   0.50    &   3.47    $\pm$   0.38    &   -10.17  $\pm$   0.15    &   47.65   $\pm$   0.38    \\
130831A $^a$    &   0.479   &   1.79    $\pm$   0.11    &   52.20   $\pm$   0.04    &   50.55   $\pm$   0.51    &   4.99    $\pm$   0.07    &   -12.05  $\pm$   0.14    &   44.72   $\pm$   0.07    \\
130606A $^a$    &   5.913   &   1.87    $\pm$   0.11    &   53.85   $\pm$   0.22    &   52.55   $\pm$   0.58    &   4.15    $\pm$   0.43    &   -11.44  $\pm$   0.28    &   47.88   $\pm$   0.43    \\
130514A $^a$    &   3.600   &   2.06    $\pm$   0.17    &   53.96   $\pm$   0.05    &   52.70   $\pm$   0.54    &   3.82    $\pm$   0.76    &   -11.32  $\pm$   0.69    &   47.60   $\pm$   0.76    \\
130505A $^a$    &   2.270   &   1.92    $\pm$   0.05    &   54.55   $\pm$   0.23    &   52.74   $\pm$   0.54    &   4.53    $\pm$   0.21    &   -10.59  $\pm$   0.05    &   47.80   $\pm$   0.21    \\
130418A $^c$    &   1.218   &   1.69    $\pm$   0.18    &   52.45   $\pm$   0.20    &   51.24   $\pm$   0.49    &   2.96    $\pm$   0.35    &   -9.81   $\pm$   0.24    &   47.90   $\pm$   0.35    \\
121211A $^a$    &   1.023   &   2.07    $\pm$   0.11    &   52.32   $\pm$   0.78    &   51.47   $\pm$   0.37    &   4.48    $\pm$   0.33    &   -11.68  $\pm$   0.17    &   45.88   $\pm$   0.33    \\
121128A $^a$    &   2.200   &   1.98    $\pm$   0.09    &   53.89   $\pm$   0.52    &   51.64   $\pm$   0.53    &   3.17    $\pm$   0.14    &   -10.06  $\pm$   0.09    &   48.32   $\pm$   0.14    \\
121027A $^a$    &   1.773   &   2.37    $\pm$   0.09    &   52.82   $\pm$   0.13    &   53.01   $\pm$   0.37    &   5.12    $\pm$   0.12    &   -12.04  $\pm$   0.04    &   46.16   $\pm$   0.12    \\
121024A $^a$    &   2.298   &   2.01    $\pm$   0.12    &   53.03   $\pm$   0.62    &   51.49   $\pm$   0.58    &   4.44    $\pm$   0.56    &   -11.77  $\pm$   0.40    &   46.66   $\pm$   0.56    \\
120729A $^o$    &   0.800   &   1.88    $\pm$   0.12    &   52.42   $\pm$   0.22    &   50.62   $\pm$   0.50    &   3.88    $\pm$   0.12    &   -11.27  $\pm$   0.02    &   46.03   $\pm$   0.12    \\
120404A $^a$    &   2.876   &   1.90    $\pm$   0.12    &   53.05   $\pm$   0.11    &   51.30   $\pm$   0.48    &   3.52    $\pm$   0.37    &   -10.92  $\pm$   0.32    &   47.71   $\pm$   0.37    \\
120327A $^a$    &   2.810   &   1.76    $\pm$   0.15    &   53.56   $\pm$   0.14    &   51.74   $\pm$   0.62    &   3.49    $\pm$   0.27    &   -10.50  $\pm$   0.04    &   48.05   $\pm$   0.27    \\
111228A $^a$    &   0.716   &   2.04    $\pm$   0.07    &   52.73   $\pm$   0.11    &   51.25   $\pm$   0.42    &   3.84    $\pm$   0.13    &   -10.67  $\pm$   0.17    &   46.50   $\pm$   0.13    \\
111123A $^a$    &   3.152   &   2.56    $\pm$   0.17    &   53.83   $\pm$   0.12    &   52.51   $\pm$   0.46    &   4.57    $\pm$   0.35    &   -12.08  $\pm$   0.08    &   46.84   $\pm$   0.35    \\
111008A $^a$    &   5.000   &   1.94    $\pm$   0.07    &   53.93   $\pm$   0.07    &   52.41   $\pm$   0.47    &   3.47    $\pm$   0.24    &   -10.46  $\pm$   0.10    &   48.75   $\pm$   0.24    \\
110801A $^a$    &   1.858   &   2.05    $\pm$   0.09    &   53.22   $\pm$   0.14    &   52.10   $\pm$   0.43    &   4.06    $\pm$   0.39    &   -11.41  $\pm$   0.18    &   46.80   $\pm$   0.39    \\
110213A $^a$    &   1.460   &   1.96    $\pm$   0.05    &   53.14   $\pm$   0.18    &   51.79   $\pm$   0.55    &   3.32    $\pm$   0.41    &   -9.61   $\pm$   0.02    &   48.32   $\pm$   0.41    \\
100906A $^a$    &   1.727   &   2.03    $\pm$   0.08    &   53.59   $\pm$   0.04    &   52.14   $\pm$   0.38    &   3.94    $\pm$   0.10    &   -10.77  $\pm$   0.17    &   47.36   $\pm$   0.10    \\
100814A $^a$    &   1.440   &   1.89    $\pm$   0.04    &   53.59   $\pm$   0.12    &   52.02   $\pm$   0.52    &   4.55    $\pm$   0.11    &   -10.98  $\pm$   0.20    &   46.93   $\pm$   0.11    \\
100704A $^a$    &   3.600   &   2.12    $\pm$   0.09    &   53.80   $\pm$   0.08    &   52.61   $\pm$   0.48    &   4.10    $\pm$   0.28    &   -11.15  $\pm$   0.24    &   47.79   $\pm$   0.28    \\
100621A $^a$    &   0.542   &   2.30    $\pm$   0.11    &   52.83   $\pm$   0.03    &   51.52   $\pm$   0.29    &   3.67    $\pm$   0.41    &   -10.47  $\pm$   0.26    &   46.38   $\pm$   0.41    \\
100615A $^a$    &   1.398   &   2.38    $\pm$   0.16    &   53.02   $\pm$   0.05    &   51.61   $\pm$   0.54    &   4.29    $\pm$   0.18    &   -10.95  $\pm$   0.09    &   46.97   $\pm$   0.18    \\
100425A $^a$    &   1.755   &   2.17    $\pm$   0.18    &   52.43   $\pm$   1.18    &   50.94   $\pm$   0.51    &   4.52    $\pm$   0.70    &   -12.36  $\pm$   0.39    &   45.81   $\pm$   0.70    \\
100418A $^a$    &   0.624   &   2.27    $\pm$   0.35    &   51.16   $\pm$   0.35    &   50.22   $\pm$   0.59    &   4.91    $\pm$   0.28    &   -12.11  $\pm$   0.06    &   44.90   $\pm$   0.28    \\
100413A $^o$    &   3.900   &   1.96    $\pm$   0.11    &   54.20   $\pm$   0.18    &   52.37   $\pm$   0.60    &   3.76    $\pm$   0.42    &   -10.59  $\pm$   0.27    &   48.37   $\pm$   0.42    \\
091020  $^a$    &   1.710   &   2.09    $\pm$   0.07    &   53.26   $\pm$   0.18    &   51.56   $\pm$   0.54    &   3.90    $\pm$   0.20    &   -10.95  $\pm$   0.00    &   47.18   $\pm$   0.20    \\
090530  $^a$    &   1.266   &   2.04    $\pm$   0.13    &   52.45   $\pm$   0.46    &   50.76   $\pm$   0.63    &   4.68    $\pm$   0.58    &   -12.08  $\pm$   0.32    &   45.71   $\pm$   0.58    \\
090516A $^a$    &   4.109   &   2.09    $\pm$   0.07    &   54.04   $\pm$   0.08    &   52.59   $\pm$   0.46    &   4.22    $\pm$   0.10    &   -11.24  $\pm$   0.14    &   47.83   $\pm$   0.10    \\
090418A $^a$    &   1.608   &   2.03    $\pm$   0.09    &   53.36   $\pm$   0.21    &   51.45   $\pm$   0.55    &   3.44    $\pm$   0.21    &   -10.24  $\pm$   0.02    &   47.81   $\pm$   0.21    \\
090113  $^c$    &   1.749   &   2.25    $\pm$   0.23    &   52.52   $\pm$   0.25    &   50.99   $\pm$   0.61    &   2.78    $\pm$   0.28    &   -10.23  $\pm$   0.02    &   47.94   $\pm$   0.28    \\
090102  $^c$    &   1.547   &   1.77    $\pm$   0.08    &   51.63   $\pm$   0.25    &   51.59   $\pm$   0.56    &   3.30    $\pm$   0.49    &   -9.91   $\pm$   0.44    &   48.06   $\pm$   0.49    \\
081008  $^a$    &   1.968   &   1.98    $\pm$   0.11    &   53.29   $\pm$   0.15    &   51.94   $\pm$   0.40    &   4.27    $\pm$   0.24    &   -11.41  $\pm$   0.03    &   46.85   $\pm$   0.24    \\
081007  $^a$    &   0.529   &   2.10    $\pm$   0.14    &   51.56   $\pm$   0.58    &   50.25   $\pm$   0.59    &   4.60    $\pm$   0.33    &   -11.85  $\pm$   0.10    &   45.00   $\pm$   0.33    \\
080928  $^a$    &   1.692   &   2.14    $\pm$   0.10    &   52.90   $\pm$   0.22    &   51.86   $\pm$   0.39    &   4.35    $\pm$   0.15    &   -11.64  $\pm$   0.05    &   46.48   $\pm$   0.15    \\
080906  $^a$    &   2.000   &   2.00    $\pm$   0.26    &   53.28   $\pm$   0.23    &   51.79   $\pm$   0.68    &   4.31    $\pm$   0.48    &   -11.26  $\pm$   0.28    &   47.02   $\pm$   0.48    \\
080905B $^a$    &   2.374   &   1.86    $\pm$   0.10    &   52.99   $\pm$   0.25    &   51.89   $\pm$   0.61    &   3.77    $\pm$   0.56    &   -10.39  $\pm$   0.34    &   48.03   $\pm$   0.56    \\
080810  $^a$    &   3.350   &   2.12    $\pm$   0.10    &   53.92   $\pm$   0.17    &   52.20   $\pm$   0.56    &   3.80    $\pm$   0.21    &   -10.76  $\pm$   0.01    &   48.10   $\pm$   0.21    \\
080707  $^a$    &   1.230   &   2.07    $\pm$   0.19    &   51.98   $\pm$   0.39    &   50.43   $\pm$   0.61    &   4.18    $\pm$   0.47    &   -11.89  $\pm$   0.17    &   45.88   $\pm$   0.47    \\
080607  $^a$    &   3.036   &   2.03    $\pm$   0.09    &   54.61   $\pm$   0.09    &   52.37   $\pm$   0.44    &   3.35    $\pm$   0.37    &   -10.38  $\pm$   0.16    &   48.35   $\pm$   0.37    \\
080430  $^a$    &   0.767   &   2.04    $\pm$   0.08    &   51.99   $\pm$   0.25    &   50.76   $\pm$   0.61    &   4.51    $\pm$   0.17    &   -11.58  $\pm$   0.11    &   45.67   $\pm$   0.17    \\
080310  $^a$    &   2.427   &   2.09    $\pm$   0.06    &   53.30   $\pm$   0.48    &   52.27   $\pm$   0.39    &   4.04    $\pm$   0.11    &   -11.39  $\pm$   0.15    &   47.12   $\pm$   0.11    \\
071021  $^a$    &   2.452   &   2.13    $\pm$   0.13    &   52.91   $\pm$   0.43    &   51.69   $\pm$   0.50    &   4.43    $\pm$   0.78    &   -11.87  $\pm$   0.96    &   46.65   $\pm$   0.78    \\
070810A $^c$    &   2.170   &   2.17    $\pm$   0.16    &   53.30   $\pm$   0.12    &   50.85   $\pm$   0.61    &   3.80    $\pm$   0.71    &   -11.54  $\pm$   0.66    &   46.86   $\pm$   0.71    \\
070714B $^a$    &   0.920   &   2.07    $\pm$   0.15    &   52.30   $\pm$   0.71    &   50.37   $\pm$   0.45    &   3.42    $\pm$   0.27    &   -11.10  $\pm$   0.09    &   46.35   $\pm$   0.27    \\
070529  $^a$    &   2.500   &   1.98    $\pm$   0.17    &   53.51   $\pm$   0.48    &   51.16   $\pm$   0.69    &   3.42    $\pm$   0.54    &   -10.92  $\pm$   0.35    &   47.59   $\pm$   0.54    \\
070318  $^a$    &   0.836   &   1.97    $\pm$   0.10    &   52.69   $\pm$   0.27    &   50.97   $\pm$   0.46    &   5.43    $\pm$   0.19    &   -12.60  $\pm$   0.03    &   44.75   $\pm$   0.19    \\
070306  $^a$    &   1.497   &   1.94    $\pm$   0.07    &   53.22   $\pm$   0.23    &   51.72   $\pm$   0.50    &   4.28    $\pm$   0.16    &   -10.77  $\pm$   0.14    &   47.19   $\pm$   0.16    \\
070208  $^c$    &   1.165   &   2.20    $\pm$   0.19    &   51.80   $\pm$   0.39    &   50.35   $\pm$   0.58    &   3.12    $\pm$   0.25    &   -10.96  $\pm$   0.09    &   46.75   $\pm$   0.25    \\
070129  $^a$    &   2.338   &   2.28    $\pm$   0.12    &   53.18   $\pm$   0.17    &   52.23   $\pm$   0.45    &   4.41    $\pm$   0.22    &   -11.75  $\pm$   0.22    &   46.76   $\pm$   0.22    \\
070110  $^a$    &   2.352   &   2.09    $\pm$   0.06    &   53.06   $\pm$   0.28    &   51.80   $\pm$   0.57    &   3.54    $\pm$   0.27    &   -10.62  $\pm$   0.12    &   47.85   $\pm$   0.27    \\
061222A $^a$    &   2.088   &   1.93    $\pm$   0.06    &   53.90   $\pm$   0.12    &   52.14   $\pm$   0.58    &   4.83    $\pm$   0.27    &   -11.41  $\pm$   0.13    &   46.90   $\pm$   0.27    \\
061121  $^a$    &   1.314   &   1.90    $\pm$   0.06    &   53.77   $\pm$   0.09    &   52.04   $\pm$   0.40    &   3.05    $\pm$   0.20    &   -9.82   $\pm$   0.14    &   48.00   $\pm$   0.20    \\
061021  $^a$    &   0.346   &   1.99    $\pm$   0.06    &   52.23   $\pm$   0.24    &   50.32   $\pm$   0.52    &   4.61    $\pm$   0.37    &   -11.47  $\pm$   0.15    &   44.95   $\pm$   0.37    \\
060814  $^a$    &   0.840   &   2.12    $\pm$   0.07    &   53.34   $\pm$   0.09    &   51.36   $\pm$   0.42    &   3.81    $\pm$   0.36    &   -10.92  $\pm$   0.05    &   46.43   $\pm$   0.36    \\
060729  $^a$    &   0.540   &   2.02    $\pm$   0.04    &   52.01   $\pm$   0.35    &   51.37   $\pm$   0.45    &   4.49    $\pm$   0.06    &   -10.79  $\pm$   0.20    &   46.09   $\pm$   0.06    \\
060719  $^a$    &   1.532   &   2.57    $\pm$   0.15    &   52.56   $\pm$   0.11    &   50.91   $\pm$   0.58    &   4.23    $\pm$   0.61    &   -11.76  $\pm$   0.41    &   46.27   $\pm$   0.61    \\
060714  $^a$    &   2.710   &   2.04    $\pm$   0.11    &   53.26   $\pm$   0.07    &   51.86   $\pm$   0.51    &   3.67    $\pm$   0.35    &   -11.18  $\pm$   0.05    &   47.44   $\pm$   0.35    \\
060614  $^a$    &   0.130   &   1.90    $\pm$   0.08    &   51.50   $\pm$   0.04    &   50.76   $\pm$   0.25    &   4.51    $\pm$   0.14    &   -11.39  $\pm$   0.12    &   44.09   $\pm$   0.14    \\
060607A $^a$    &   3.082   &   1.61    $\pm$   0.05    &   53.50   $\pm$   0.20    &   52.30   $\pm$   0.52    &   4.11    $\pm$   0.05    &   -10.41  $\pm$   0.13    &   48.17   $\pm$   0.05    \\
060605  $^a$    &   3.800   &   2.02    $\pm$   0.09    &   52.99   $\pm$   0.44    &   51.57   $\pm$   0.61    &   3.77    $\pm$   0.32    &   -11.00  $\pm$   0.06    &   47.96   $\pm$   0.32    \\
060604  $^a$    &   2.136   &   2.17    $\pm$   0.12    &   52.24   $\pm$   0.58    &   51.58   $\pm$   0.51    &   4.40    $\pm$   0.20    &   -11.82  $\pm$   0.11    &   46.57   $\pm$   0.20    \\
060526  $^a$    &   3.210   &   1.91    $\pm$   0.12    &   53.04   $\pm$   0.35    &   52.38   $\pm$   0.46    &   4.33    $\pm$   0.26    &   -11.03  $\pm$   0.17    &   47.71   $\pm$   0.26    \\
060502A $^a$    &   1.510   &   2.03    $\pm$   0.12    &   53.04   $\pm$   0.24    &   51.31   $\pm$   0.63    &   4.44    $\pm$   0.36    &   -11.50  $\pm$   0.06    &   46.48   $\pm$   0.36    \\
060210  $^a$    &   3.910   &   2.08    $\pm$   0.05    &   54.06   $\pm$   0.19    &   52.59   $\pm$   0.52    &   4.46    $\pm$   0.17    &   -11.12  $\pm$   0.07    &   47.90   $\pm$   0.17    \\
\hline
\end{longtable}
} }

\end{document}